\documentclass[12pt,a4paper,twoside]{article}
\usepackage{setspace}
\usepackage{graphicx,hyperref,bm,url,microtype,color,bbm}
\usepackage{amsfonts,amssymb,amsthm,amsmath,MnSymbol}
\AtEndDocument{\refstepcounter{equation}\label{finalequation}} % referencias cruzadas
\usepackage{pgfplots}
\usepackage[english]{babel}
\usepackage{natbib}
\usepackage{comment} % Para usar en entorno comment, que ignora todo lo que hay dentro
\usepackage{tikz}
\usepackage{float} % If for some reason you really want some particular figure to be placed “HERE”, and not where LaTeX wants to put it, then use the [H] option of the “float” package which basically turns the floating figure into a regular non-float.

\usepackage{multirow}

\usepackage{caption}
\usepackage{subcaption}

\usepackage[a4paper,text={16.5cm,23.5cm},centering]{geometry}
\usepackage[compact,small]{titlesec}
\usepackage[utf8]{inputenc}
\usepackage{enumitem} % Para decir de manera f\'{a}cil en el entorno enumerate el contador que queremos: 1), (1), a), (a), i), (i), etc.

\def \d{\text{\rm d}}

\def \E{\text{\rm E}}
\def \P{\text{\rm P}}

\def \B{\mathbb{B}}
\newcommand{\R}{\ensuremath{\mathbb{R}}}

\newcommand{\I}{\ensuremath{\mathcal{I}}}

\DeclareMathOperator*{\Var}{Var}

\DeclareMathOperator{\sgn}{sgn}

\definecolor{violet}{rgb}{0.7,0,0.6}
\theoremstyle{definition}

\newtheorem{example}{Example}%[section]
\newtheorem{remark}{Remark}

\newtheorem{assumption}{Assumption}

\theoremstyle{theorem}

\newtheorem{theorem}{Theorem}
\newtheorem{proposition}{Proposition}
\newtheorem{corollary}{Corollary}

\newenvironment{keywords}
    {\vspace*{3mm}
    {\noindent{}\textit{Keywords\/:}}
        \nopagebreak\small}
        {}

\title{\textbf{\Large Tests for almost stochastic dominance}}
\author{Amparo Ba\'illo$^1$, Javier C\'{a}rcamo$^{2,}$\footnote{Corresponding author: Javier C\'{a}rcamo.
E-mail Address: javier.carcamo@ehu.eus}\, and Carlos Mora-Corral$^{1,3}$ \\
{\small{$^{1}$ Departamento de Matem\'{a}ticas, Universidad Aut\'{o}noma de Madrid, 28049 Madrid (SPAIN)}}\\
{\small{$^{2}$ Departamento de Matem\'{a}ticas, Universidad del Pa\'{\i}s Vasco, Aptdo. 644, 48080 Bilbao (SPAIN)}} \\
{\small{$^{3}$  Instituto de Ciencias Matem\'{a}ticas, CSIC-UAM-UC3M-UCM, 28049 Madrid (SPAIN)}}
}
\date{}

\begin{document}
%-------------------------------------------------------------------

\maketitle

%-------------------
\begin{abstract}
\noindent
We introduce a 2-dimensional stochastic dominance (2DSD) index to characterize both strict and almost stochastic dominance. Based on this index, we derive an estimator for the \textit{minimum violation ratio} (MVR), also known as the \textit{critical parameter},
of the almost stochastic ordering condition between two variables.
%\textcolor{blue}{, considered the  in almost dominance}.
We determine the asymptotic properties of the empirical 2DSD index and MVR for the most frequently used stochastic orders.
%\textcolor{blue}{: first dominance or usual stochastic order; second stochastic dominance; convex-type orderings; and the Lorenz order; among others}.
We also provide conditions under which the bootstrap estimators of these quantities are strongly consistent. As an application, we develop consistent bootstrap testing procedures for almost stochastic dominance.
%In particular, we consider a test with the alternative hypothesis of almost stochastic dominance.
The performance of the tests is checked via simulations and the analysis of real data.
%a bootstrap testing procedure for the alternative hypothesis of almost stochastic dominance. The 2DSD index can also be used to obtain a consistent point estimator and a confidence interval for the \textit{violation ratio} of the stochastic ordering. The testing methodology is simple and it can applied to the most frequently used stochastic orders: first dominance or usual stochastic order; second stochastic dominance or convex ordering; and the Lorenz order; among others. The performance of the test is checked with simulations and the analysis of real data sets.
\end{abstract}
%------------------

%\newpage

%-----------------
%\begin{MSC}

%\end{MSC}
%--------------------------------------------------------------------------
%\newpage

%%%%%%%%%%%%%%%%%%%%%%%%%%%%%%%%%%%%%%%%%%%%%%%%%%%%%%%%%%%%%%%%%%%%%%%%%%%%%%%%%%%
% SECTION: INTRODUCTION
%%%%%%%%%%%%%%%%%%%%%%%%%%%%%%%%%%%%%%%%%%%%%%%%%%%%%%%%%%%%%%%%%%%%%%%%%%%%%%%%%%%

\section{Introduction and motivation} \label{Section.Introduction}

In this section we present the main ideas of this work, as well as the general context in which they are included and the motivation of our proposals.

\subsubsection*{Stochastic dominance}%. Background and motivation} \color{black}

Stochastic orders, also known as \textit{stochastic dominance (SD) rules} in the economic literature, are partial order relations in the set of probability measures; see \cite{Shaked-Shanthikumar-2006} and \cite{Levy-2016}.
%are sourcebooks on this subject in probability  and decision theory under uncertainty, respectively.
As SD rules allow comparing and ranking random elements (according to some specific criterion), they are useful in many different disciplines.
%For instance, SD constraints are often included in mathematical optimization problems.
%Stochastic orders are also fundamental to define several concepts of aging, which are of interest in the analysis of the reliability of systems.
Economics is one of the areas in which stochastic orders are especially relevant and appear more frequently. There is a large body of literature on this topic in fields such as inequality analysis, actuarial science, portfolio insurance, operations management, inventory or financial optimization.
%For instance, they are used to compare the underlying risks under two possible decisions or income inequality among different societies.
%Stochastic orders usually perform global comparisons of the distributions, in the sense that the differences between the variables are evaluated through a specific \textit{target function} instead of simple summarizing indices. The selection of the target function depends on the problem. For instance, a natural choice of the target function is the (integrated) distribution function of the variables, which leads to first (second) stochastic dominance rule (respectively) in risk analysis or the (generalized) Lorenz curve in the context of inequality and social welfare.

An advantage of (global) stochastic comparisons between two variables over (partial) comparisons  through summary statistics is that, if two distributions are stochastically ordered---with respect to a certain relation---we can derive many important consequences regarding the underlying distributions. Typically, SD implies an ordering among many features or summary measures of the underlying distributions simultaneously. For instance, if the Lorenz ordering holds, then several inequality measures (such as the Gini index) of the involved variables are ordered in the same way; see \cite{Arnold-Sarabia-2018}. Moreover, many dominance rules are in fact \textit{integral stochastic orders} (see \cite{Muller-1997}), that is, they are defined by comparing expected values of functions (in a certain class) of the variables. In this way, SD is linked to expected utility theory in the context of decision theory under uncertainty; see \cite{Fishburn-1984}.

Given the practical importance of checking, based on observed data, whether it is (statistically) reasonable to assume that two random variables, \(X_1\) and \(X_2\), are ordered, in the literature
there are various hypothesis tests of the type
\begin{equation} \label{Dtest}
\begin{cases}
H_0: & X_1 \preceq X_2,\\
H_1: &  X_1 \npreceq X_2, \\
\end{cases}
\end{equation}
where `$\preceq$' denotes the stochastic order of interest; see, e.g., \cite{Anderson-1996}, \cite{Barrett-Donald-2003}, \cite{Zheng-2002}, \cite{Berrendero-Carcamo-2011}, \cite{Barrett-Donald-Bhattacharya-2014} and \cite{Sun-Beare-2021}.% \textcolor{blue}{Creo que conviene quitar alguna(s) referencias(s) de aqu\'{\i}.}  \textcolor{violet}{Pues yo creo que no.}

\subsubsection*{Almost stochastic dominance and minimum violation ratio}

SD rules are partial orders and, hence, not every pair of distributions can be ranked.
%; see \cite{Davies-Hoy-1995}.
To overcome this important drawback, \cite{Leshno-Levy-2002} introduced the notion of \textit{almost stochastic dominance} (ASD), which refers to the situation where strict SD %\textcolor{magenta}{of either distribution of the pair over the other distribution}
does not hold but the region where the dominance rule is violated is relatively small. In risk analysis, ASD describes the situation where ``most'' decision makers would prefer one uncertain prospect over another one, which is equivalent to eliminating some extreme or pathological utility functions from the ordering condition.
One advantage of relaxing the strict dominance criterion by its almost counterpart is that the latter is a more flexible concept and allows ordering more pairs of distributions.
%One advantage of relaxing the strict dominance criterion by its almost counterpart is that, usually under mild conditions, the latter always holds between two random variables.
A second advantage of ASD is that its definition (see Section~\ref{Section.Stochastic-order}) intrinsically depends on the degree to which the strict stochastic dominance, say $X_1 \preceq X_2$, is unfulfilled.
This quantity, usually called \textit{violation ratio} and denoted by the parameter $\epsilon$ (with $0\leq\epsilon\leq 1$), has an intuitive interpretation: small values of $\epsilon$ mean that the variables are very close to being ordered; \cite{Levy-2012}. Conversely, the larger the \(\epsilon\), the more distant of strict dominance is the relationship between the variables.

%In Section~\ref{Section.Stochastic-order} we give the precise definition of ASD with violation ratio \(\epsilon\) employed in this paper. }
In what follows of this introduction we state the questions of interest at the core of this work using only the above intuitive definition of ASD. For this purpose, let us denote by `$\preceq^\epsilon$' the ASD rule with parameter $\epsilon >0$. We will see in Section~\ref{Section.Stochastic-order} that, if $X_1 \preceq^{\epsilon_1} X_2$ holds for a certain $\epsilon_1>0$, then $X_1 \preceq^{\epsilon_2} X_2$, for all $\epsilon_2>\epsilon_1$. Therefore, in this context it is important to determine the \textit{minimum violation ratio} (MVR), that is, the smallest possible value of $\epsilon$ for which the two variables are almost ordered, given by
\begin{equation}\label{MVR}
\epsilon_0=\inf\{ \epsilon\in[0,1] : X_1 \preceq^\epsilon X_2\}.
\end{equation}
In the literature there is yet no clear distinction between the concept of the MVR $\epsilon_0$ and that of a single $\epsilon$ for which ASD holds.
%In general, the quantities $\epsilon$ and the MVR $\epsilon_0$ in \eqref{MVR} could be different.
The value of $\epsilon$
can be preselected by the practitioner as the acceptable degree of non-fulfillment in the SD condition. For instance, using solely the sample information, we might be interested in checking whether the ASD holds with $\epsilon=0.05$ when actually the (unknown) population MVR is $\epsilon_0=0.01$.
%However, in the literature there is yet no clear distinction between these two concepts.

%The violation ratio $\epsilon$ in \eqref{ASD} represents the degree of noncompliance in the order relationship that the supervisor is willing to accept. In this context,

%\cite{Leshno-Levy-2002} mainly considered the usual stochastic order in which the (survival) distribution functions have to be pointwise ordered. In this case, the ASD amounts to violating this condition within a set of small length so that the proportion of the ``area violation'' is less or equal to a small number $\epsilon>0$; see the precise definition below. The parameter $\epsilon$ quantifies the \textit{unfulfilment degree} in the stochastic ordering definition; small values of $\epsilon$ mean that the variables are very close to being ordered. Almost second stochastic dominance, as well as various generalizations (almost $N$th-degree stochastic dominance), is considered in \cite{Tzeng-Huang-Shih-2013}, \cite{Tsetlin-et-al-2015}, and \cite{Chang-et-al-2019}. We refer the reader to \cite[Chapter 7]{Levy-2016} for a general overview. Later on, \cite{Zheng2018} extends this idea to the Lorenz ordering. From the point of view of utility theory, ASD is equivalent to some extreme utility functions, called ``pathological'' in \cite{Levy-2016} being eliminated from the ordering condition. In the context of risk analysis, this means that one variable is preferred than the other one for \textit{almost} all decision makers. Hence, only those deciders with extreme preferences are eliminated with respect to ASD.

\subsubsection*{The topic of this paper}

In practice, after observing samples from two variables, the statement that two distributions are ordered cannot be proved statistically, as one would desire.
That is, interchanging the roles of the null and alternative hypotheses in \eqref{Dtest} results in an ill-posed problem.
%This happens because a test with null hypothesis ``two variables are  \textit{not} ordered" and alternative ``the variables are ordered" is usually ill-posed.
Indeed, given a pair of ordered distributions, we can generally find pairs of non-ordered distributions arbitrarily close to the initial ones and hence the null and alternative hypotheses are indistinguishable; see \cite{Ermakov-2017}. However, ASD provides a more flexible setting, since the collection of pairs of almost ordered distributions is much larger than the set of pairs of strictly ordered variables. Thus, we can implement statistical tests of the form
\begin{equation} \label{ASDtest}
\begin{cases}
H_0: & X_1 \npreceq^\epsilon X_2,\\
H_1: & X_1 \preceq^\epsilon X_2,
\end{cases}
\end{equation}
where the ASD assumption is placed in the \textit{alternative hypothesis}.
Observe that rejection of the null hypothesis in \eqref{ASDtest} means that there is statistical evidence that the two variables are almost ordered, while, in the usual tests of type \eqref{Dtest}, non-rejection of \(H_0\) only means there is not enough evidence against it. This difference between the conclusions of \eqref{Dtest} and \eqref{ASDtest} makes the ASD test relevant in many applications where the aim is to establish the (almost) ordering between the variables; see \cite{Huang-Kan-Tzeng-Wang-2021}.
%Actually, the test \eqref{ASDtest} can be seen as a natural step before estimating the minimum violation ratio.
%In other words, rejection of the null hypothesis in \eqref{ASDtest} means that there is a solid statistical evidence that the two variables are almost ordered, while in the usual tests of type \eqref{Dtest} we can only assert that it is not unreasonable to assume that the variables are ordered. We believe that this difference between the final conclusions of \eqref{Dtest} and \eqref{ASDtest} makes the ASD test potentially relevant in many applications in which we have real data and the aim is to determine the (almost) ordering between the variables.

As noted by \cite{Huang-Kan-Tzeng-Wang-2021},
%(the estimation of)
the MVR \(\epsilon_0\) in \eqref{MVR}
%, also called \textit{critical parameter} by these authors,
is a key point in the analysis of ASD relations. Further, the definition of the MVR (and its distinction from the  parameter $\epsilon$) allows us to write the problems that we consider here in a clear and concise way. For instance, given a fixed value of $\epsilon>0$, test \eqref{ASDtest} is equivalent to $H_0: \epsilon_0>\epsilon$ versus $H_1:\epsilon_0\le \epsilon$. In this work we are interested in the following hypothesis tests:
\begin{equation}\label{Test.MVR}
\text{(a)}\quad
\begin{cases}
H_0: & \epsilon_0 \ge \epsilon,\\
H_1: & \epsilon_0 < \epsilon,
\end{cases}
\qquad\quad \text{(b)}\quad
\begin{cases}
H_0: & \epsilon_0 \le \epsilon,\\
H_1: & \epsilon_0 > \epsilon,
\end{cases}
\qquad\quad \text{(c)}\quad
\begin{cases}
H_0: & \epsilon_0 = \epsilon,\\
H_1: &\epsilon_0 \ne \epsilon.
\end{cases}
\end{equation}
Observe that test (a) in \eqref{Test.MVR} essentially coincides with the ASD test in \eqref{ASDtest} with
a standard and better-posed formulation of null and alternative. Test (b) corresponds to $H_0:\ X_1 \preceq^\epsilon X_2$ versus $H_1:\ X_1 \npreceq^\epsilon X_2$, a weaker version of the usual SD test in \eqref{Dtest}.
%; this second test aims at deciding if it is reasonable to accept the $\epsilon$-ASD.
The last test in (c) amounts to obtaining a confidence interval for the MVR.

%an \(\epsilon\) for which ASD holds and the value \(\epsilon^*\).
%intermingled

\subsubsection*{Our contributions}

In this work we present a general framework to treat, in a unified way, the problems that we consider for several stochastic orders simultaneously, as well as their ``almost'' versions. Our main contributions are the following:

\begin{enumerate}[label={(\arabic*)}, topsep=0mm, partopsep=0mm, itemsep=0mm, parsep=0mm, align=left, labelwidth=*, leftmargin=0 mm]
  \item [(1)]  We define the concepts of $s$-SD and $s$-ASD that include as particular cases the common examples of stochastic orders in the literature: the usual stochastic order (first-degree SD); the second SD and the stop-loss order (convex and concave-type orderings); the Lorenz order; as well as higher degree relationships together with their ``almost'' counterparts.

  \item [(2)] We introduce a 2-dimensional stochastic dominance (2DSD) index, defined for pairs of variables, that characterizes both proper SD and ASD. The 2DSD index has two main purposes: on the one hand, it plays a central role in the estimation of the MVR \(\epsilon_0\) in \eqref{MVR} and, hence, in the construction of suitable rejection regions of the ASD tests in \eqref{Test.MVR}. On the other hand, the index serves to estimate and visualize the MVR of the stochastic dominance condition, as well as to evaluate the ``distance'' between the distributions in a simple way through a 2-dimensional plot. %In addition, the graphical representation of the index over time makes it possible to follow the evolution of two populations at different moments. This type of pairwise visual comparisons might be valuable in those disciplines (such as economics) in which there are often annual data available or quantifications at different points in time (income, salaries, etc.).

  \item [(3)] We carry out a complete asymptotic study of the plug-in estimators of the 2DSD index and the MVR $\epsilon_0$ in \eqref{MVR}. We show their strong consistency and find the limiting distributions in the usual examples. The almost sure consistency of the standard bootstrap estimators of these quantities is also established.

   \item [(4)] We propose a (consistent) bootstrap rejection region for the tests of ASD in \eqref{Test.MVR} for the most used stochastic orders.
%In this work we focus on  tests for almost dominance such as \eqref{ASDtest} for different stochastic orders (the most used in practice).For each specific stochastic relation, the test statistic is based on the 2DSD index.
The practical behaviour of our procedure is evaluated through some simulations and real data examples.
%As we explain in Section \ref{Section.index}, a unidimensional index (as test statistic) does not have the required properties to carry out this testing procedure.
\end{enumerate}

%The 2DSD indices that we introduce have two main purposes: on the one hand, they play a central role in the construction of the rejection region of the tests \eqref{ASDtest}, one of the main goals of this work. On the other hand, these indices have an independent interest since they allow practitioners to visualize the degree of fulfillment of the stochastic dominance condition, as well as evaluating the ``distance'' between the distributions in a simple way through a 2-dimensional plot.

\subsubsection*{Organization of this paper}

%The paper is organized as follows.
%The structure of this work is the following.
In Section \ref{Section.Stochastic-order} we define the notions of $s$-SD and $s$-ASD.
%In Section \ref{Section.Stochastic-order} we present the basic definitions that we use throughout this work: $s$-Stochastic Dominance and $\epsilon$-Almost $s$-Stochastic Dominance.
In Section \ref{Section.index} we introduce the 2DSD index and enumerate its main properties, in particular how it characterizes both strict SD and ASD.
Section \ref{Section-Asymptotics} introduces the plug-in estimators of the 2DSD index and the MVR. Under suitable conditions, we prove the strong consistency and determine the asymptotic distributions of the normalized estimators. We also show the (a.s.) consistency of the corresponding bootstrap estimators. In particular, we provide a consistent resampling scheme to approximate the asymptotic distribution of any statistic defined through the $\mathrm{L}^1$-norm.
%The integrability and smoothness assumptions that we have to impose to obtain the asymptotic results ultimately depend on the target function $s$ that we are using. However, the general lines allow us to adapt the results to different $s$ and the proofs are reduced to showing the convergence of the underlying stochastic processes in the space $\mathrm{L}^1$.
In Section \ref{Section.ASDtest} we propose bootstrap rejection regions, based on the estimated MVR, for the ASD tests in \eqref{Test.MVR}. The procedure is intuitive and fast to compute for the usual orders even for moderately large data sets. The performance of the tests is checked through simulations.
%In Section \ref{Section.ASDtest} we implement the test \eqref{ASDtest} using 2DSD indices. We opt for a bootstrap procedure (on the indices) to find the test rejection regions. Section \ref{Section.ALDincome} is devoted to analyse various real data set with the propose methodology.
In Section~\ref{Section.RealData} we analyse two real data sets with the proposed methodology. %Specifically, we discuss the gender wage gap in Spain with the first order almost dominance and the almost Lorenz dominance between income datasets.
The detailed proofs of the results are collected in the
online Appendix.

%%%%%%%%%%%%%%%%%%%%%%%%%%%%%%%%%%%%%%%%%%%%%%%%%%%%%%%%%%%%%%%%%%%%%%%%%%%%%%%%%%%
% SECTION: Stochastic orders and bidimensional indices
%%%%%%%%%%%%%%%%%%%%%%%%%%%%%%%%%%%%%%%%%%%%%%%%%%%%%%%%%%%%%%%%%%%%%%%%%%%%%%%%%%%

\section{$s$-Stochastic Dominance and $s$-Almost Stochastic Dominance} \label{Section.Stochastic-order}

%Examples of stochastic orders: In terms of quantile functions? Merece la pena poner un ejemplo para las relaciones superiores?

%Stochastic orders usually perform global comparisons of the distributions, in the sense that the differences between the variables are evaluated through a specific \textit{target function} \(s\) instead of simple summarizing indices. The function \(s\) measures the feature that we want to compare with the SD rule and characterizes the stochastic order. The choice of this function eventually depends on the problem of interest.

A stochastic order relationship between two random elements $X_1$ and $X_2$ is typically established by the pointwise comparison of a real function $s:D_s\to\R$ in the two populations, in the sense that we can define the comparison rule:
\begin{equation} \label{SDs}
X_1 \preceq_{s} X_2 \quad \mbox{if and only if} \quad s_1(t)\ge s_2(t), \quad \mbox{for all} \; t \in D_s,
\end{equation}
where $s_1$ and $s_2$ denote the function $s$ in populations \(X_1\) and \(X_2\), respectively, and \(D_s\) is a suitable set, usually the domain of $s$ or the support of the variables.
%(typically, \(\R\) or an interval).
If \eqref{SDs} holds, we say that $X_2$ \textit{stochastically dominates $X_1$ with respect to $s$}, and denote it by $X_2$ $s$-SD $X_1$.

We refer to  $s$ as the \textit{target function} of the stochastic ordering as it represents the characteristic that we want to compare in both populations: survival time, distribution of wealth, salaries, accumulated risk, total size, variability, concentration, dependency, etc. The function \(s\) corresponding to the population \(X\) usually depends on the probability distribution of \(X\) in a way determined by the specific stochastic order under consideration.
The framework of $s$-SD encompasses many well-known SD rules.
%Finally, \(D_s\) is a suitable set, usually the domain of $s$ or the support of the variables. Below we list some examples that are included within the framework of $s$-SD.

\begin{example}[First stochastic dominance or the usual stochastic order] \label{Example1stOrderSD}
When the target function is the distribution function of a random variable $X$,
$s(t)=F(t)=\P(X\le t)$, $t\in \R$, the order $\preceq_s$ in \eqref{SDs} is the \textit{usual stochastic order} or \textit{first (stochastic) dominance rule}. This ordering ranks distributions according to their global size.
%Let us consider
%$$s(t)=F(t)=\P(X\le t),\quad  t\in \R,$$
%the distribution function of a random variable $X$. The ordering $\preceq_s$ in \eqref{SDs} is well-known in many scientific fields as the \textit{usual stochastic order}, denoted by $\le_{\text{st}}$. In economics it is called the \textit{first (stochastic) dominance rule}, usually denoted by $\le_1$. For instance, if $X_1$ and $X_2$ represent two risks, clearly we prefer $X_1$ than $X_2$ whenever $X_1\le_1 X_2$ since the probability that $X_1$ exceeds the value $t$ is always less than or equal to the corresponding one for $X_2$, for all $t\in \R$.
\end{example}

\begin{example}[Second stochastic dominance or the increasing concave order] \label{Example2ndOrderSD}
For the integrated distribution function of $X$,
%\begin{equation} \label{IntegratedDistribFn}
\begin{equation}\label{eq.s.second.dominance}
s(t)=\int_{-\infty}^t F(x)\,\d x,\quad t\in \R,
\end{equation}
the corresponding order $\preceq_s$ is the \textit{second stochastic dominance rule} or  \textit{increasing concave order}.  %(see \cite[p. 182]{Shaked-Shanthikumar-2006}).
Higher orders ($N$th-degree stochastic dominance) can be analogously defined by successively integrating the distribution functions. The second-degree SD takes into account both size and variability of the distributions.

%We consider
%$$s(t)=\int_{-\infty}^t F(x)\,\d x,\quad t\in \R,$$
%the integrated distribution function of $X$. The corresponding ordering $\preceq_s$ is the \textit{second stochastic dominance rule}, denoted by $\le_2$. In the literature of probability theory (and other applications) this relation is called the  \textit{increasing concave order}, $\le_{\text{icv}}$, as it is the integral stochastic order generated by increasing and concave functions; see \cite[p. 182]{Shaked-Shanthikumar-2006}.
%Higher orders (as $N$th-degree stochastic dominance) can be analogously defined by successively integrating the distribution functions of the two variables.
\end{example}

\begin{example}[Stop-loss order or the increasing convex order] \label{ExampleStopLoss}
When $
s(t)=%\pi_X(t)=
\int_{t}^\infty (1-F(x))\,\d x$,  $t\in  \R$, the integrated survival function of $X$, the ordering $\preceq_s$ is the
\textit{stop-loss order} in actuarial sciences (see \cite{Denuit et al}), the \textit{risk-seeking stochastic dominance rule} in risk theory (see \cite{Levy-2016}) and the \textit{increasing convex order} in probability theory (see \cite{Shaked-Shanthikumar-2006}).
As in Example~\ref{Example2ndOrderSD}, higher degree stochastic relationships are defined by integrating recursively.
%Let us consider
%$$s(t)=\pi_X(t)=\int_{t}^\infty (1-F(x))\,\d x,\quad t\in  \R,$$
%the integrated survival function of $X$.  The ordering $\preceq_s$ is called the
%\textit{stop-loss order}, $\le_{\text{sl}}$, in actuarial sciences and $\pi_X(t)$ is the \textit{stop-loss transform}; see \cite{Denuit et al}. This relationship is also called the \textit{risk-seeking stochastic dominance rule} in risk theory; see \cite{Levy-2016}.  The function $\pi_X(t)=\E(X-t)_+$, where `$\E$' denotes mathematical expectation and $(a)_+$ indicates the positive part of the number $a$, can be considered as the net premium of a stop-loss reinsurance contract; see \cite{Muller-Stoyan-2002}. Further, in probability theory $\preceq_s$ is known as the  \textit{increasing convex order}, $\le_{\text{icx}}$, as it is the integral order generated by increasing and convex functions; see \cite[Section 4]{Shaked-Shanthikumar-2006}.
%Thanks to the duality between convex and concave functions, it holds that $X\le_{\text{icx}} [\le_{\text{icv}}] Y$ if and only if $-X\le_{\text{icv}} [\le_{\text{icx}}] - Y$. Therefore, the stop loss order and the second  dominance rule are closely related. As in the previous example, higher degree stochastic relationships can be defined analogously by integrating recursively.
\end{example}

\begin{example}[Lorenz ordering and inverse stochastic dominance] \label{ExampleLorenz}
Given a positive and integrable random variable $X$ with expectation $\E(X)=\mu >0$, its \textit{Lorenz curve} is
\begin{equation} \label{Lorenz-curve}
\ell(t)=\frac{1}{\mu}\int_0^t F^{-1}(x)\, \d x,\quad 0\le t \le 1,
\end{equation}
where, for $0<x<1$, $F^{-1}(x)=\inf\{ y \ge 0 : F(y)\ge x \}$ is the \textit{quantile function} of $X$.
%, that is, the generalized inverse of $F$.
If $s=\ell$ in \eqref{Lorenz-curve}, then $\preceq_s$ is the \textit{Lorenz ordering}.
%, denoted here by $\le_{\text{L}}$, an important instrument to compare distributions according to inequality.
Repeatedly integrating the Lorenz curve \eqref{Lorenz-curve} we obtain \textit{inverse stochastic dominance} rules (see \cite{Muliere-Scarsini-1989} or \cite{Cal-Carcamo-2010}), used to rank intersecting Lorenz curves.
%Let $X$ be a positive and integrable random variable with mean $\E(X)=\mu >0$. The Lorenz curve (of $X$) is given by
%\begin{equation}
%\label{Lorenz-curve}
%\ell(t)=\frac{1}{\mu}\int_0^t F^{-1}(x)\, \d x,\quad 0\le t \le 1,
%\end{equation}
%where $F^{-1}(x)=\inf\{ y \ge 0 : F(y)\ge x \}$ ($0<x<1$) is the \textit{quantile function} of $X$, that is, the generalized inverse of $F$.
%If $s(t)=\ell(t)$ ($t\in (0,1)$), then $\preceq_s$  is the so-called \textit{Lorenz ordering}, denoted here by $\le_{\text{L}}$, an important instrument to compare distributions according to inequality.
%Again, repeatedly integrating the Lorenz curve \eqref{Lorenz-curve} (on the interval $[0,t]$) we obtain the so-called \textit{inverse stochastic dominance} rules; see \cite{Muliere-Scarsini-1989}, \cite{Aaberge-2008}, \cite{Cal-Carcamo-2010} and the references therein. These higher orderings allow ranking intersecting Lorenz curves.
\end{example}

%\section{$s$-Almost Stochastic Dominance ($s$-ASD)} \label{Section.ASD}

The definition of SD in terms of a function \(s\) becomes more flexible when the condition `\(s_1(t)\ge s_2(t)\)' in (\ref{SDs}) is allowed to fail in a ``small'' subset of \(t\)'s. From now on we assume that any pair \((s_1,s_2)\) fulfils that the difference \(s_1-s_2\in \mathrm{L}^1\), the  space of functions $s:D_s \to \R$ endowed with the usual $\mathrm{L}^1$-norm,
%\begin{equation*}
\(\Vert s\Vert =\int_{D_s} |s(t)|\, \d t\) for \(s\in \mathrm{L}^1\).
%\end{equation*}
Generalizing previous definitions of ASD (\cite{Leshno-Levy-2002}, \cite{Tsetlin-et-al-2015}, and \cite{Zheng2018}), for $\epsilon \ge 0$, we consider the following ASD rule:
%The definition of stochastic dominance in (\ref{SDs}) in terms of a function \(s\) is more flexible and adapts to more settings when the condition \(s_1(t)\ge s_2(t)\) is allowed to fail in ``a small'' subset of \(t\)'s. From now on we assume that any pair \((s_1,s_2)\) fulfils that the difference \(s_1-s_2\in \mathrm{L}^1\), where $\mathrm{L}^1=\mathrm{L}^1(D_s)$ is the Banach space of (equivalence classes of measurable) functions $s:D_s \to \R$ endowed with the usual $\mathrm{L}^1$-norm denoted in the following as $\Vert \cdot \Vert$. That is,
%\begin{equation*}
%\Vert s\Vert =\int_{D_s} |s(t)|\, \d t,\quad \text{for }s\in \mathrm{L}^1.
%\end{equation*}
%Generalizing previous definitions of ASD (\cite{Leshno-Levy-2002}, \cite{Tzeng-Huang-Shih-2013}, \cite{Tsetlin-et-al-2015}, \cite{Chang-et-al-2019}, and \cite{Zheng2018}), for $\epsilon \ge 0$, we propose that
\begin{equation} \label{ASD}
X_1\preceq_s^\epsilon X_2 \quad \mbox{if and only if} \quad  \int_{\{s_1<s_2\}} (s_2(t)-s_1(t))\, \d t \le \epsilon \Vert s_1-s_2\Vert.
\end{equation}
If \eqref{ASD} holds, we say that $X_2$ \textit{$\epsilon$-almost $s$-stochastically dominates} $X_1$, written \(X_2\) $s$-ASD($\epsilon$) \(X_1\).
\cite{Leshno-Levy-2002} defined ASD for the first and second-degree SD.
Subsequently, \cite{Tsetlin-et-al-2015} extended this idea to the $N$th-degree SD (\(N>2\)) and \cite{Zheng2018} considered the Lorenz ordering.

The degree of noncompliance of the stochastic ordering is quantified by the parameter $\epsilon$, called the {\em relative area violation} by \cite{Levy-2012} and the {\em violation ratio} of the ordering condition by \cite{Huang-Kan-Tzeng-Wang-2021}. Observe that when the functions $s_1$ and $s_2$ are smooth enough (right-continuous, for instance), a value \(\epsilon=0\) is equivalent to proper SD as in (\ref{SDs}). This is fulfilled for the common choices of \(s\) in the most frequently used stochastic orders; see Examples \ref{Example1stOrderSD}--\ref{ExampleLorenz}. Note that it always holds that $X_1 \preceq_s^\epsilon X_2$ for $\epsilon\ge 1$. Thus, only values $\epsilon<1$ make sense. Further, it is always satisfied that $X_1\preceq_s^{0.5} X_2$ or $X_2\preceq_s^{0.5} X_1$.
%Therefore, it is usually considered that $\epsilon\in [0,0.5]$ since we can choose the ordering between variables according to the lowest degree of non-compliance with the stochastic ordering condition.

\begin{remark}\label{rk:s1s2L1}
The requirement \(s_1-s_2\in \mathrm{L}^1\) usually amounts to an integrability condition on the  variables \(X_1\) and \(X_2\). For instance, if \(F_j\) denotes the distribution function of \(X_j\), \(j=1,2\), for the first SD (Example~\ref{Example1stOrderSD}) we impose that $F_1-F_2 \in \mathrm{L}^1$, which is fulfilled when both $\E(X_1)$ and $\E(X_2)$ are finite. In the case of inverse SD (Example~\ref{ExampleLorenz}), once the Lorenz curves of the variables exist (which happens whenever the variables have finite means), ASD makes perfect sense.

However, for the second SD (Example \ref{Example2ndOrderSD}), the function
$$s_1(t)-s_2(t)  =  \int_{-\infty}^{t} (F_1(x)-F_2(x))\, \d x, \quad t\in \R,$$
is \textit{not} integrable in general. The same happens for the stop-loss order of Example \ref{ExampleStopLoss}. In fact, since
%$$
\(s_1(t)-s_2(t) \to \E  X_2 - \E X_1 \) as \(t\to\infty\),
%$$
we have that $s_1-s_2\notin \mathrm{L}^1$ if $\E X_1 \ne \E X_2$ and the support of the variables is unbounded. To overcome this limitation, in the literature it is commonly assumed that $X_1$ and $X_2$ take values in a reference bounded interval; see \cite{Tzeng-Huang-Shih-2013}, \cite{Tsetlin-et-al-2015} and \cite{Chang-et-al-2019}.
%If such interval is denoted by \([a,b]\), then \(s:D_s=[a,b]\to\R\) is defined as
%\begin{equation} \label{s_order2}
%s(t) = \int_a^t F(x)\,\d x, \quad  t\in [a,b].
%\end{equation}
If the assumption of compact support is too demanding or unrealistic,
%The assumption of compact support (of the underlying variables) in second SD rules (Examples \ref{Example2ndOrderSD}-\ref{ExampleStopLoss}) seems too demanding and perhaps unrealistic considering the usual distributions used in many applied disciplines to model the observed phenomena (such as lifetimes, income or salaries, risks, etc.). In fact, in many of these situations the distributions that appear are often continuous, with unbounded support (usually $(0,\infty)$ or $\R$) and even heavy-tailed (as is the case when modeling income or wealth). However, if we are willing
then, for \(s_1-s_2\in \mathrm{L}^1\) to hold, it suffices to assume that the variables have equal means ($\E X_1=\E X_2$) and finite second moment ($\E X_1^2$,  $\E X_2^2<\infty$).
%we can remove the bounded support condition.
As a matter of fact, under these two assumptions and for the second SD,
% for $s_1(t)-s_2(t)  =  \int_{-\infty}^{t} (F_1(x)-F_2(x))\, \d x$,
by Fubini's theorem it can be checked that $\Vert s_1-s_2\Vert\le (\E X_1^2+ \E X_2^2)/2$.
A further possibility to guarantee integrability of \(s_1-s_2\) is to trim the tails of the distributions.
%\begin{align*}
%\Vert s_1-s_2\Vert   &  =  \int_{-\infty}^{0}  \left|  \int_{\infty}^{t} (F_1(x)-F_2(x))\, \d x   \right| \d t +  \int_{0}^{+\infty}  \left| \int_{t}^{\infty} (F_1(x)-F_2(x))\, \d x   \right| \d t  \\
%   & \le  \int_{0}^{+\infty}  x (1-F_1(x))\, \d x -  \int_{-\infty}^{x} x F_1(x)\, \d x  \\
%   & \quad +  \int_{0}^{+\infty}  x (1-F_2(x))\, \d x -  \int_{-\infty}^{x} x F_2(x)\, \d x  \\
%   & = \frac{1}{2}(\E X_1^2+ \E X_2^2).
%\end{align*}
%Therefore, in this situation ASD is well-defined even for unbounded pairs of distributions.
%Regarding the condition of equality of means in the second SD (Example \ref{Example2ndOrderSD}), we observe that $X_1\le_2 X_2$ if and only if there exists a random variable $Y$ with $\E Y = \E X_2$ such that $X_1\le_1 Y \le_2 X_2$; see \cite[Theorem  4.A.6]{Shaked-Shanthikumar-2006}. Hence, $X_1\le_2 X_2$ can be seen as a difference in size (between $X_1$ and $Y$) plus a difference in variability (between $Y$ and $X_2$).
% {\color{blue}
% For instance, if the variables have a normal distribution, $X_i\sim \mathcal{N}(\mu_i,\sigma_i)$ ($i=1,2$), then in Examples \ref{Example1stOrderSD} and \ref{ExampleLorenz} $s_1-s_2\in L^1$ as the variables are integrable. However, in Examples \ref{Example2ndOrderSD} and \ref{ExampleStopLoss} $s_1-s_2\in L^1$ if and only if $\mu_1=\mu_2$ because $X_1$ and $X_2$ have finite second moment.
% }
\end{remark}

\section{2-Dimensional Stochastic Dominance (2DSD) indices} \label{Section.index}

To carry out the tests
% test \eqref{ASDtest}, as well as those
in \eqref{Test.MVR}, we propose to use a bidimensional index to characterize both strict SD and ASD defined in (\ref{SDs}) and (\ref{ASD}), respectively.
%As a consequence, we derive an intuitive procedure to test the significance of \(\epsilon\)-ASD as in (\ref{ASDtest}).
There are several reasons behind the need for a two-dimensional characterization of SD and ASD instead of using a simpler one-dimensional statistic.
First, it is convenient that any unidimensional index intending to measure relative dominance (of one variable with respect to another one) takes values on a reference (fixed) bounded symmetric interval.
Assume that some normalization has allowed this interval to be $[-1,1]$, in such a way that the extremes of the interval correspond to the strict SD cases, e.g., the index is 1 when \(X_1\preceq_s X_2\) and $-1$ when \(X_2\preceq_s X_1\).
Secondly, the index should be continuous in the sense that if there are two random sequences \((X_{1,n_1})\), \( (X_{2,n_2})\) converging to $X_1$ and $X_2$, as \(n_1,n_2\to\infty\), respectively, in some suitable mode of stochastic convergence, then the values of the dominance index between \(X_{1,n_1}\) and \(X_{2,n_2}\) should converge to the index between \(X_1\) and \(X_2\).
This continuity property is essential in practice, when using random samples from the populations to construct an empirical counterpart of the index.
However, these two requirements are incompatible for a one-dimensional index.
For example, we can think of a sequence \(X_{1,n}\) converging to \(X\) as \(n\to\infty\) such that \(X_{1,n}\preceq_s X\) for all \(n\). The sequence of indices corresponding to \((X_{1,n},X)\) would be \(1\) for all \(n\), but the index of \((X,X)\) would be naturally \(0\).

%For example, we can think of \textcolor{blue}{two sequences \(X_{1,n}\) and \(X_{2,n}\) converging to \(X\) as \(n\to\infty\) such that \(X_{1,n}\preceq_s X_{2,n}\) for all \(n\). The sequence of indices corresponding to \((X_{1,n},X_{2,n})\)} would be \(1\) for all \(n\), but the index of \((X,X)\) would be naturally \(0\).

However, it is possible to construct an easy-to-handle, two-dimensional index characterizing both the strict SD \eqref{SDs}, as well as the $\epsilon$-ASD (\ref{ASD}). Indeed, our following proposal satisfies, among other desirable properties, the two requirements described before: symmetry and continuity.
%; see Proposition \ref{Proposition.I.properties.1}  below.
We define the {\em 2DSD  (2-dimensional Stochastic Dominance) index}, associated to the ordering $\preccurlyeq_s$ in \eqref{SDs}, as
\begin{equation} \label{2DSD-index}
\mathcal{I}_s (X_1,X_2) = \mathcal{I} (s_1,s_2) = \left( \int_{D_s} (s_1-s_2),\int_{D_s}|s_1-s_2| \right) .
\end{equation}
%If \(s_1,s_2\in \mathrm{L}^1\) and $s_1,s_2\ge 0$,
%(as in the case of Lorenz curves in Example \ref{ExampleLorenz}, for instance)
%then the 2DSD index in \eqref{2DSD-index} admits the more %compact expression
%\[
%\mathcal{I}_s (X_1,X_2) = (\Vert s_1\Vert-\Vert %s_2\Vert,\Vert s_1-s_2\Vert).
%\]
This definition is inspired by the inequality index introduced in \cite{Baillo-Carcamo-Mora-2022}. The first component of the index \eqref{2DSD-index} provides a signed value on the fulfillment
%(in one way or another)
of the $s$-SD between the two variables. The second component is a measure of the ``distance'' between the two distributions in terms of the function $s$.
%the variables measured as the $\mathrm{L}^1$-norm between the target functions in the two populations. This second component has different interpretations depending on the underlying function $s$; see the examples below.
In Sections~\ref{Section-Asymptotics} and~\ref{Section.ASDtest} we use the new index \eqref{2DSD-index} to estimate the MVR in \eqref{MVR} and to define suitable critical regions for the ASD tests \eqref{Test.MVR}.

For the examples of target functions \(s\) considered in Section~\ref{Section.Stochastic-order}, we derive the expression of the 2DSD index.

%\begin{example}[First stochastic dominance index]\label{Example1stOrderSDI}
{\bf Example~\ref{Example1stOrderSD} {\rm (First stochastic dominance index)}.}
Let $F_1$ and $F_2$ be the distribution functions of two integrable variables $X_1$ and $X_2$, respectively. The 2DSD index for the first stochastic order is given by
\begin{equation}\label{2DSD-1}
%\mathcal{I}_F (X_1,X_2) =
\mathcal{I}_1 (X_1,X_2) = \left( \mu_2-\mu_1 , d_W(X_1,X_2)\right),
\end{equation}
where \(\mu_1\) and \(\mu_2\) are the expectations of \(X_1\) and \(X_2\), respectively, and
\begin{equation}\label{Wasserstein-distance}
d_W(X_1,X_2) = \int_{-\infty}^{+\infty} |F_1-F_2|= \int_{0}^{1} |F_1^{-1}-F_2^{-1}|
\end{equation}
is the celebrated $\mathrm{L}^1$-\textit{Wasserstein distance}; see \cite{Villani-2009}.
%By the Kantorovich–Rubinstein theorem, we have that
%\begin{equation}\label{Zolotarev-metrics}
%d_W(X_1,X_2)=\sup \left\{|\E f(X_1)-\E f(X_2)| :  f\in\mathcal{F}_1 \right\},
%\end{equation}
%where $\mathcal{F}_1$ is the class of real-valued functions $f$ on $\R$ having first derivative $f'$ a.e., and such that $|f'|\le 1$ a.e. In particular, $d_W(X_1,X_2)$ is an integral probability metric (see \cite{Rachev-2013}) that gives us the maximum (absolute) difference between  $X_1$ and $X_2$ in expected value within the class of smooth functions $\mathcal{F}_1$ . Therefore, $\mathcal{I}_1 (X_1,X_2)$ in \eqref{2DSD-1} simultaneously measures the relative size of the variables (quantify in the first component as the difference of their means) and the Wasserstein distance between them (second component of the index).
%\end{example}

%\begin{example}[Second stochastic dominance index]\label{Example2ndOrderSDI}
{\bf Example~\ref{Example2ndOrderSD} {\rm (Second stochastic dominance index)}.}
%We consider the stochastic order generated by $s(t)=\int_{-\infty}^t F(x)\,\d x$; see Example \ref{Example2ndOrderSD}. As we have commented in the previous section, in this example ASD does not make sense unless the distributions have compact support or equal means. In the latter case, that is,
Assuming that $\E X_1=\E X_2$ and that $\E(X_1^2)$ and $\E(X_2^2)$ are finite, we have
\begin{equation}\label{2DSD-2}
\mathcal{I}_2 (X_1,X_2) = \left( (\Var (X_1)-\Var(X_2))/2, d_2(X_1,X_2)\right),
\end{equation}
where $\Var(X)$ stands for the variance of $X$ and $d_2(X_1,X_2)$ is the \textit{Zolotarev (ideal) metric} of order $2$; see \cite{Rachev-2013}.
Under the same assumptions, it turns out that the 2DSD index for the stop-loss order (Example \ref{ExampleStopLoss}) has the same expression as \eqref{2DSD-2}.
Likewise, according to Remark \ref{rk:s1s2L1}, assumptions $\E X_1=\E X_2$ and $\E(X_1^2) , \E(X_2^2) < \infty$ can be dispensed with if the variables $X_1$ and $X_2$ have compact support.
 %We omit the details.
%\textcolor{green}{given by
%\begin{equation}\label{Zolotarev-metrics}
%d_2(X_1,X_2)=\sup \left\{|\E f(X_1)-\E f(X_2)| :  f\in\mathcal{F}_2 \right\},
%\end{equation}
%where $\mathcal{F}_2$ is the class of real-valued functions $f$ on $\R$ having $2$-nd derivative $f^{(2)}$ a.e. and such that $|f^{(2)}|\le 1$ a.e. Hence, $\mathcal{I}_2 (X_1,X_2)$ measures the relative variability of the %variables in the first component (half the difference between their variances) and the Zolotarev distance between them (second component).} \textcolor{blue}{?`Es necesario dar la f\'{o}rmula de la distancia, etc, etc? Yo quitar\'{\i}a todo lo que est\'{a} escrito en verde.}

%\textcolor{blue}{Tambi\'{e}n quitar\'{\i}a el sgte p\'{a}rrafo (no es necesario dar tanto detalle):}
%We note that if $X_1$ and $X_2$ do not have equal means, but they take values in a bounded interval $[a,b]$, the first component of the 2SDS index is given by
%$$ \int_{a}^{b}  \int_{a}^{t} (F_1(x) -F_2(x))\, \d x\, \d t =  (\E(X_1^2)-\E(X_2^2))/2 + b (\E X_2 -\E X_1).$$
%\end{example}

%\textcolor{blue}{{\bf Example~\ref{ExampleStopLoss} {\rm (Stop-loss order)}.}}

%\begin{example}[Lorenz dominance index]\label{ExampleLorenzI}
{\bf Example~\ref{ExampleLorenz} {\rm (Lorenz dominance index)}.}
Let $X_1$ and $X_2$ be positive and integrable random variables with Lorenz curves $\ell_1$ and  $\ell_2$, respectively. The 2DSD index associated to the Lorenz order is given by %\textcolor{blue}{Seleccionar una notaci\'{o}n \(\mathcal{I}_\ell\) u otra \(\mathcal{I}_\text{L}\), pero no poner las dos.}
\begin{equation}\label{2DSD-Lorenz}
\mathcal{I}_\ell (X_1,X_2)% = \mathcal{I}_\text{L} (\ell_1,\ell_2)
= \left( (G(\ell_2)-G(\ell_1))/2, \|\ell_1-\ell_2\|\right),
\end{equation}
where \( G(\ell)=  1-2 \|  \ell  \| \) is the \textit{Gini index} of the variable $X$ with Lorenz curve $\ell$. This was essentially the index introduced by \cite{Baillo-Carcamo-Mora-2022}.
%The Gini index is maybe the most important inequality measure and has many interesting interpretations and representations; see \citet[Chapter 2]{Yitzhaki-Schechtman-2012}. %, where more than a dozen alternative ways of spelling Gini are presented.
%Therefore, the index in \eqref{2DSD-Lorenz} quantifies relative inequality (expressed as the difference between the Gini indices of the variables) in the first component and the $\mathrm{L}^1$-distance between the Lorenz curves  in the second component.
%\end{example}

The following result collects the main properties of the 2DSD index $\mathcal{I}_s$ defined in \eqref{2DSD-index}.

\begin{proposition}\label{Proposition.I.properties.1}
Let $X_1$ and $X_2$ be two variables with target functions $s_1$ and $s_2$, respectively, and assume that $s_1-s_2\in \mathrm{L}^1(D_s)$. We have that:
\begin{enumerate}[label={(\alph*)}, topsep=0mm, partopsep=0mm, itemsep=0mm, parsep=0mm, align=left, labelwidth=*, leftmargin=0 mm]
\item\label{item:Property.Range} \textsl{Range:} The image of the index $\mathcal{I}_s (X_1, X_2)$ is contained in the ``triangle" %defined by
%\begin{equation*}%\label{eq:Delta}
\(\Delta= \{ (x,y)\in \R\times [0,\infty) : |x|\le y \}\).
%\end{equation*}

\item\label{item:Property.Symmetry} \textsl{Symmetry:} If $\mathcal{I}_s (X_1, X_2) = (x,y) $, then $\mathcal{I}_s (X_2, X_1) = (-x,y) $.

\item\label{item:Property.SD} \textsl{Characterization of $s$-SD:}  If $s_1$ and $s_2$ are right-continuous, then
\begin{enumerate}[label={(\arabic*)}, topsep=0mm, partopsep=0mm, labelindent=7mm, itemsep=0mm, parsep=0mm, align=left, labelwidth=*, leftmargin=0 mm]
\item $X_1 \preceq_s X_2$ if and only if $\mathcal{I}_s (X_1,X_2)\in L_1= \{ (x,x) : x\ge 0  \}$.
\item $X_2 \preceq_s X_1$ if and only if $\mathcal{I}_s (X_1,X_2)\in L_2= \{ (-x,x) : x\ge 0  \}$.
\end{enumerate}
Therefore, strict $s$-SD between $X_1$ and $X_2$ holds if and only if $\mathcal{I}_s (X_1,X_2)$ lies on the frontier of $\Delta$. In particular, $\mathcal{I}_s (X_1,X_2)=(0,0)$ if and only if $X_1=_s X_2$, that is, $X_1 \preceq_s X_2$ and $X_2 \preceq_s X_1$ simultaneously.

\item\label{item:Property.ASD}  \textsl{Characterization of $s$-ASD:} For $0\le \epsilon < 1/2$, we have that
\begin{enumerate}[label={(\arabic*)}, topsep=0mm, partopsep=0mm, labelindent=7mm, itemsep=3mm, parsep=0mm, align=left, labelwidth=*, leftmargin=0 mm]
    \item $X_1 \preceq^\epsilon_s X_2$ if and only if $\mathcal{I}_s (X_1,X_2)\in R_{1,\epsilon} = \{ (x,y)\in \Delta : y \le x/(1-2 \epsilon) \}$.
    \item $X_2 \preceq^\epsilon_s X_1$ if and only if $\mathcal{I}_s (X_1,X_2)\in R_{2,\epsilon} = \{ (x,y)\in \Delta : y \le -x/(1-2 \epsilon) \}$.
  \end{enumerate}

  In particular, if $s_1\ne s_2$, 
%  In particular, if $\mathcal{I}_s (X_1, X_2) = (x,y)$ with $y>0$,
the MVR in \eqref{MVR} can be written as a function of the index components as
%  \begin{equation}\label{MVR.Characterization}
%    \epsilon_0 = \frac{1}{2}\left(   1- \frac{x}{y}    \right).
%  \end{equation}
\begin{equation}\label{MVR.Characterization}
\epsilon_0 = \frac{1}{2} \left( 1- \frac{\int_{D_s} (s_1-s_2)}{\int_{D_s} |s_1-s_2|} \right).
\end{equation}

\item\label{item:Property.Continuity}  \textsl{Continuity:} If $\{s_{1,n_1}\}_{n_1\ge 1}$ and $\{s_{2,n_2}\}_{n_2\ge 1}$ are sequences such that $s_{1,n_1}-s_{2,n_2}\to s_1-s_2$ in $\mathrm{L}^1$ as $n_1,n_2\to\infty$, then
         $\I_s(s_{1,n_1},s_{2,n_2})\to \I_s(s_1,s_2)$.
\end{enumerate}
\end{proposition}

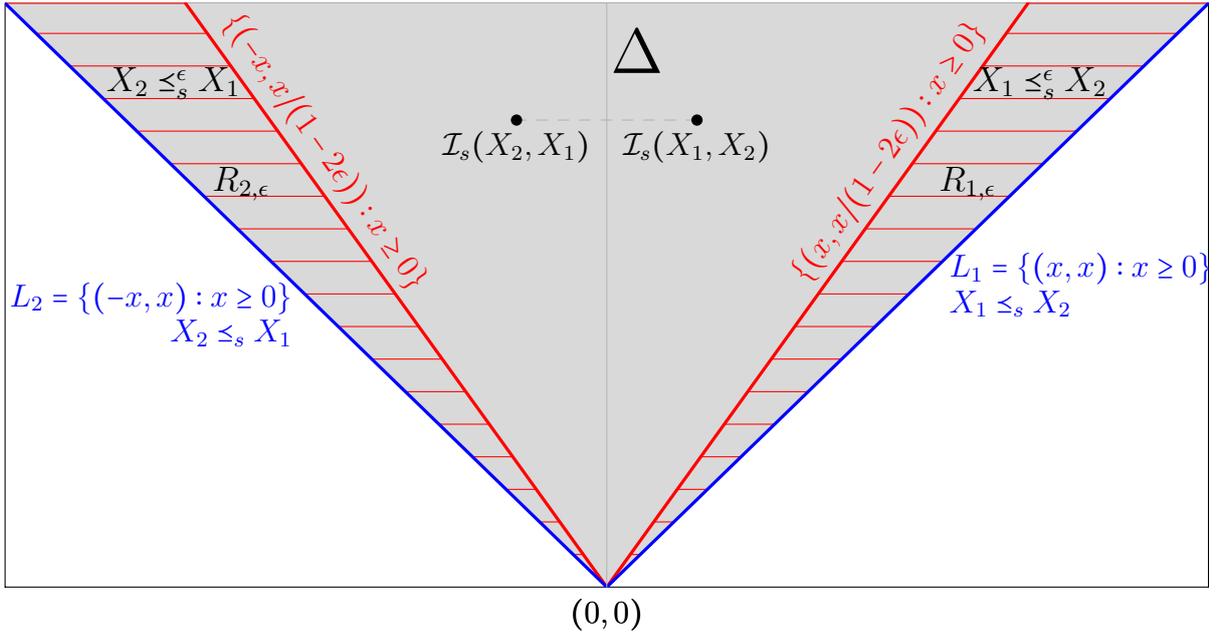
\begin{figure}\centering{
\begin{tikzpicture}[x=1pt,y=1pt,scale=0.85]
\definecolor{fillColor}{RGB}{255,255,255}
\path[use as bounding box,fill=fillColor,fill opacity=0.00] (0,0) rectangle (542.02,289.08);
\begin{scope}
\path[clip] (  6.00, 24.00) rectangle (536.02,283.08);
\definecolor{drawColor}{RGB}{0,0,0}

\path[draw=drawColor,line width= 0.4pt,line join=round,line cap=round] (536.02, 24.00) --
	(542.02, 24.00);
\end{scope}
\begin{scope}
\path[clip] (  0.00,  0.00) rectangle (542.02,289.08);
\definecolor{drawColor}{RGB}{0,0,0}

\path[draw=drawColor,line width= 0.4pt,line join=round,line cap=round] (  6.00, 24.00) --
	(536.02, 24.00) --
	(536.02,283.08) --
	(  6.00,283.08) --
	(  6.00, 24.00);
\end{scope}
\begin{scope}
\path[clip] (  0.00,  0.00) rectangle (542.02,289.08);
\definecolor{drawColor}{RGB}{0,0,0}

\node[text=drawColor,anchor=base,inner sep=0pt, outer sep=0pt, scale=  1.00] at (271.01,  8.40) {$(0,0)$};

\node[text=drawColor,anchor=base,inner sep=0pt, outer sep=0pt, scale=  1.00] at (271.01,  8.40) {$(0,0)$};
\end{scope}
\begin{scope}
\path[clip] (  6.00, 24.00) rectangle (536.02,283.08);
\definecolor{drawColor}{gray}{0.85}
\definecolor{fillColor}{gray}{0.85}

\path[draw=drawColor,line width= 0.4pt,line join=round,line cap=round,fill=fillColor] (271.01, 24.00) --
	(536.02,283.08) --
	(536.02,283.08) --
	(271.01,283.08) --
	cycle;

\path[draw=drawColor,line width= 0.4pt,line join=round,line cap=round,fill=fillColor] (  6.00,283.08) --
	(271.01, 24.00) --
	(271.01,283.08) --
	(  6.00,283.08) --
	cycle;
\definecolor{drawColor}{RGB}{255,0,0}

\path[draw=drawColor,line width= 0.4pt,line join=round,line cap=round] (271.01, 24.00) -- (271.01, 24.00);

\path[draw=drawColor,line width= 0.4pt,line join=round,line cap=round] (281.36, 38.45) -- (285.80, 38.45);

\path[draw=drawColor,line width= 0.4pt,line join=round,line cap=round] (291.71, 52.91) -- (300.58, 52.91);

\path[draw=drawColor,line width= 0.4pt,line join=round,line cap=round] (302.06, 67.36) -- (315.37, 67.36);

\path[draw=drawColor,line width= 0.4pt,line join=round,line cap=round] (312.41, 81.82) -- (330.15, 81.82);

\path[draw=drawColor,line width= 0.4pt,line join=round,line cap=round] (322.76, 96.27) -- (344.94, 96.27);

\path[draw=drawColor,line width= 0.4pt,line join=round,line cap=round] (333.11,110.72) -- (359.72,110.72);

\path[draw=drawColor,line width= 0.4pt,line join=round,line cap=round] (343.46,125.18) -- (374.51,125.18);

\path[draw=drawColor,line width= 0.4pt,line join=round,line cap=round] (353.81,139.63) -- (389.29,139.63);

\path[draw=drawColor,line width= 0.4pt,line join=round,line cap=round] (364.16,154.09) -- (404.08,154.09);

\path[draw=drawColor,line width= 0.4pt,line join=round,line cap=round] (374.51,168.54) -- (418.86,168.54);

\path[draw=drawColor,line width= 0.4pt,line join=round,line cap=round] (384.86,182.99) -- (433.65,182.99);

\path[draw=drawColor,line width= 0.4pt,line join=round,line cap=round] (395.21,197.45) -- (448.43,197.45);

\path[draw=drawColor,line width= 0.4pt,line join=round,line cap=round] (405.56,211.90) -- (463.22,211.90);

\path[draw=drawColor,line width= 0.4pt,line join=round,line cap=round] (415.91,226.36) -- (478.00,226.36);

\path[draw=drawColor,line width= 0.4pt,line join=round,line cap=round] (426.25,240.81) -- (492.79,240.81);

\path[draw=drawColor,line width= 0.4pt,line join=round,line cap=round] (436.60,255.26) -- (507.57,255.26);

\path[draw=drawColor,line width= 0.4pt,line join=round,line cap=round] (446.95,269.72) -- (522.36,269.72);

\path[draw=drawColor,line width= 0.4pt,line join=round,line cap=round] (271.01, 24.00) --
	(536.02,283.08) --
	(536.02,283.08) --
	(456.52,283.08) --
	(271.01, 24.00) --
	(271.01, 24.00);

\path[draw=drawColor,line width= 0.4pt,line join=round,line cap=round] (271.01, 24.00) -- (271.01, 24.00);

\path[draw=drawColor,line width= 0.4pt,line join=round,line cap=round] (256.23, 38.45) -- (260.66, 38.45);

\path[draw=drawColor,line width= 0.4pt,line join=round,line cap=round] (241.44, 52.91) -- (250.31, 52.91);

\path[draw=drawColor,line width= 0.4pt,line join=round,line cap=round] (226.66, 67.36) -- (239.96, 67.36);

\path[draw=drawColor,line width= 0.4pt,line join=round,line cap=round] (211.87, 81.82) -- (229.61, 81.82);

\path[draw=drawColor,line width= 0.4pt,line join=round,line cap=round] (197.09, 96.27) -- (219.27, 96.27);

\path[draw=drawColor,line width= 0.4pt,line join=round,line cap=round] (182.30,110.72) -- (208.92,110.72);

\path[draw=drawColor,line width= 0.4pt,line join=round,line cap=round] (167.52,125.18) -- (198.57,125.18);

\path[draw=drawColor,line width= 0.4pt,line join=round,line cap=round] (152.73,139.63) -- (188.22,139.63);

\path[draw=drawColor,line width= 0.4pt,line join=round,line cap=round] (137.95,154.09) -- (177.87,154.09);

\path[draw=drawColor,line width= 0.4pt,line join=round,line cap=round] (123.16,168.54) -- (167.52,168.54);

\path[draw=drawColor,line width= 0.4pt,line join=round,line cap=round] (108.38,182.99) -- (157.17,182.99);

\path[draw=drawColor,line width= 0.4pt,line join=round,line cap=round] ( 93.59,197.45) -- (146.82,197.45);

\path[draw=drawColor,line width= 0.4pt,line join=round,line cap=round] ( 78.81,211.90) -- (136.47,211.90);

\path[draw=drawColor,line width= 0.4pt,line join=round,line cap=round] ( 64.02,226.36) -- (126.12,226.36);

\path[draw=drawColor,line width= 0.4pt,line join=round,line cap=round] ( 49.24,240.81) -- (115.77,240.81);

\path[draw=drawColor,line width= 0.4pt,line join=round,line cap=round] ( 34.45,255.26) -- (105.42,255.26);

\path[draw=drawColor,line width= 0.4pt,line join=round,line cap=round] ( 19.67,269.72) -- ( 95.07,269.72);

\path[draw=drawColor,line width= 0.4pt,line join=round,line cap=round] (  6.00,283.08) --
	(271.01, 24.00) --
	(271.01, 24.00) --
	( 85.50,283.08) --
	(  6.00,283.08) --
	(  6.00,283.08);

\path[draw=drawColor,line width= 1.2pt,line join=round,line cap=round] (271.01, 24.00) --
	(456.52,283.08);

\path[draw=drawColor,line width= 1.2pt,line join=round,line cap=round] (271.01, 24.00) --
	( 85.50,283.08);
\definecolor{drawColor}{RGB}{0,0,255}

\path[draw=drawColor,line width= 1.2pt,line join=round,line cap=round] (  6.00,283.08) --
	(271.01, 24.00) --
	(271.01, 24.00) --
	(536.02,283.08);
\definecolor{drawColor}{gray}{0.70}

\path[draw=drawColor,line width= 0.4pt,line join=round,line cap=round] (271.01, 24.00) -- (271.01,283.08);
\definecolor{drawColor}{RGB}{0,0,255}

\node[text=drawColor,anchor=base west,inner sep=0pt, outer sep=0pt, scale=  1.00] at (422,161.61) {$L_1 = \{(x,x):x\geq 0\}$};

\node[text=drawColor,anchor=base west,inner sep=0pt, outer sep=0pt, scale=  1.00] at (422,146.06) {$X_1 \preceq_{s} X_2$};
\definecolor{drawColor}{RGB}{0,0,0}

\node[text=drawColor,anchor=base,inner sep=0pt, outer sep=0pt, scale=  2.00] at (284.26,252.17) {$\Delta$};
\definecolor{drawColor}{RGB}{0,0,255}

\node[text=drawColor,anchor=base east,inner sep=0pt, outer sep=0pt, scale=  1.00] at (132.51,148.65) {$L_2 = \{(-x,x):x\geq 0\}$};

\node[text=drawColor,anchor=base east,inner sep=0pt, outer sep=0pt, scale=  1.00] at (132.51,133.11) {$X_2\preceq_{s} X_1$};
\definecolor{drawColor}{RGB}{255,0,0}

\node[text=drawColor,rotate= 55.01,anchor=base,inner sep=0pt, outer sep=0pt, scale=  1.00] at (398.22,215.06) {$\{(x,x/(1-2\epsilon)):x\geq 0\}$};

\node[text=drawColor,rotate=304.99,anchor=base,inner sep=0pt, outer sep=0pt, scale=  1.00] at (143.81,215.06) {$\{(-x,x/(1-2\epsilon)):x\geq 0\}$};
\definecolor{drawColor}{RGB}{0,0,0}

\node[text=drawColor,anchor=base,inner sep=0pt, outer sep=0pt, scale=  1.10] at (461.82,244.06) {$X_1\preceq_{s}^{\epsilon} X_2$};

\node[text=drawColor,anchor=base,inner sep=0pt, outer sep=0pt, scale=  1.10] at (430,200) {$R_{1,\epsilon}$};

\node[text=drawColor,anchor=base,inner sep=0pt, outer sep=0pt, scale=  1.10] at ( 80.20,244.06) {$X_2\preceq_{s}^{\epsilon} X_1$};

\node[text=drawColor,anchor=base,inner sep=0pt, outer sep=0pt, scale=  1.10] at (110,200) {$R_{2,\epsilon}$};

\definecolor{drawColor}{gray}{0.70}

\path[draw=drawColor,line width= 0.4pt,dash pattern=on 4pt off 4pt ,line join=round,line cap=round] (310.76,231.26) --
	(231.26,231.26);
\definecolor{drawColor}{RGB}{0,0,0}
\definecolor{fillColor}{RGB}{0,0,0}

\path[draw=drawColor,line width= 0.4pt,line join=round,line cap=round,fill=fillColor] (310.76,231.26) circle (  2.25);

\node[text=drawColor,anchor=base,inner sep=0pt, outer sep=0pt, scale=  1.00] at (310.76,215.81) {$\mathcal{I}_{s} (X_{1},X_{2})$};

\path[draw=drawColor,line width= 0.4pt,line join=round,line cap=round,fill=fillColor] (231.26,231.26) circle (  2.25);

\node[text=drawColor,anchor=base,inner sep=0pt, outer sep=0pt, scale=  1.00] at (231.26,215.81) {$\mathcal{I}_{s} (X_{2},X_{1})$};
\end{scope}
\end{tikzpicture}
}
\caption{Summary of the properties of the 2DSD index.}
\label{fi:propiedades}
\end{figure}

%A graphical representation of the previous properties is displayed in Figure~\ref{fi:propiedades}.
Figure~\ref{fi:propiedades} provides a graphical outline of the previous properties. Parts \emph{\ref{item:Property.SD}} and \emph{\ref{item:Property.ASD}} in Proposition \ref{Proposition.I.properties.1} show that the 2DSD index \(\mathcal{I}_s (X_1,X_2)\) %in \eqref{2DSD-index}
characterizes both proper $s$-SD and $s$-ASD($\epsilon$).
Equation \eqref{MVR.Characterization} is crucial to derive a suitable test statistic for the ASD hypothesis tests \eqref{Test.MVR}.
%\begin{equation} \label{ASDtest-s}
%\begin{array}{ll}
%H_0: & X_1 \npreceq^\epsilon_s X_2,\\
%H_1: & X_1 \preceq^\epsilon_s X_2,
%\end{array}
%\end{equation}
%in which we are interested.
Property \emph{\ref{item:Property.Continuity}} provides the statistical consistency of the potential estimators of the index whenever the underlying empirical functions (playing the role of $s_{1,n_1}$ and $s_{2,n_2}$) converge in $\mathrm{L}^1$ to their population versions. This $\mathrm{L}^1$ convergence has to be established for the particular choice of the target function $s$; see Section \ref{Section-Asymptotics}.

%%%%%%%%%%%%%%%%%%%%%%%%%%%%%%%%%%%%%%%%%%%%%%%%%%%%%%%%%%%%%%%%%%%%%
% SECTION 4: ESTIMATION AND ASYMPTOTIC PROPERTIES OF THE 2DSD INDEX
%%%%%%%%%%%%%%%%%%%%%%%%%%%%%%%%%%%%%%%%%%%%%%%%%%%%%%%%%%%%%%%%%%%%%

\section{Estimation and asymptotic properties}\label{Section-Asymptotics}

In this section we carry out a deep analysis of the asymptotic properties of the natural estimators of the 2DSD index in \eqref{2DSD-index} and the associated MVR in \eqref{MVR} for the $s$-ASD. We also provide conditions that guarantee the consistency of the bootstrap estimators of these quantities. In the proofs (see the Supplemental Material) we use empirical process theory, results on differentiability of functionals and the extended functional delta method.

\subsection{Plug-in estimation of the 2DSD index and strong consistency} \label{PlugInEstimator}

First, we propose a natural estimator for the 2DSD index given in \eqref{2DSD-index}.
For $j=1,2$, we denote by $\{X_{j,i}\}_{i=1}^{n_j}$, $n_j \in\mathbb{N}$, a random sample from $X_j$. For simplicity, we assume that both samples are mutually independent. However, convergence results similar to those in this section can be obtained when we observe ``matched pairs", $ \{ (X_{1,i},X_{2,i}) \}_{i=1}^n $, drawn from a bivariate distribution $(X_1,X_2)$ with copula $C$ satisfying that its maximal correlation is strictly less than one; see \cite{Beare-2010}. As pointed out in \cite{Barrett-Donald-Bhattacharya-2014} and \cite{Sun-Beare-2021}, this second setting is more reasonable when we have one sample of individuals and two measures of the variable of interest.

%To simplify the notation, in the sequel all estimated quantities are denoted with a ``hat", and it will be implicitly understood the dependence on the corresponding sample sizes.

The estimation of the 2DSD index $\mathcal{I}_s$ depends on the selected target function $s$. In the usual practical situations, as in Examples \ref{Example1stOrderSD}--\ref{ExampleLorenz}, $s$ is expressed in terms of the distribution function, $F$, or the quantile function, \(F^{-1}\). In such cases, we propose a \textit{plug-in estimator} of $\mathcal{I}_s$ based on the empirical measures of the samples.
%Let $F_1$ and $F_2$ be the distribution functions of $X_1$ and $X_2$, respectively.
For $j=1,2$, we consider the \emph{empirical distribution functions} of the samples
\begin{equation}\label{empirical.distribution.function}
\hat{F}_j(x)=\frac{1}{n_j}\sum_{i=1}^{n_j} 1_{\{X_{j,i}\le x\}}, \quad x\in \R,
\end{equation}
where $1_A$ stands for the indicator function of the set $A$.
The corresponding \emph{empirical quantile functions} are $\hat{F}^{-1}_j(x)=\inf\{ y \ge 0 : \hat F_j (y)\ge x \}$ ($0<x<1$).
The plug-in estimator \(\hat s\) of \(s\) is obtained by replacing \(F\) (resp., \(F^{-1}\)) with its empirical counterpart $\hat{F}$ (resp., \(\hat{F}^{-1}\)). Then, the \textit{plug-in estimator of the index} $\I_s$ is the empirical 2DSD index given by
\begin{equation} \label{estimators-indices}
\hat \I(s_1,s_2) = \I (\hat s_1 ,\hat s_2 ).
\end{equation}

For instance, for the Lorenz order (Example~\ref{ExampleLorenz}) the \emph{empirical Lorenz curves} are
\begin{equation*}%\label{Lorenz-empirical}
\hat\ell_j(t)=\frac{1}{\hat \mu_j }\int_{0}^{t} \hat{F}_j^{-1} (x)\, \d x, \quad t\in[0,1],
\end{equation*}
where $\hat \mu_j = \sum_{i=1}^{n_j} X_{j,i}/n_j$ are the sample means. Then, the empirical 2DSD index is
%\begin{equation*} %\label{estimators-indices}
\(\hat \I_\ell (X_1,X_2) =  (\Vert \hat\ell_1\Vert-\Vert \hat\ell_2\Vert,\Vert \hat\ell_1-\hat\ell_2\Vert)\).
%\end{equation*}

The next proposition shows that the strong consistency of the empirical index $\hat \I_s (s_1,s_2)$ in \eqref{estimators-indices} follows from the $\mathrm{L}^1$ almost sure convergence of the estimators $\hat s_1$ and $\hat s_2$ to their population counterparts.% $s_1$ and $s_2$, respectively.

\begin{proposition}\label{Proposition:consitency}
  Let us assume that $\Vert \hat s_1 - s_1 \Vert \to 0$ and $\Vert \hat s_2 - s_2 \Vert \to 0$ a.s., as $n_1,n_2\to\infty$. Then, $\hat \I_s (s_1,s_2)\to \I_s (\hat s_1 ,\hat s_2 )$ a.s.
\end{proposition}

The next result provides, for some usual target functions, simple conditions under which the \(\hat s_j\) are $\mathrm{L}^1$ a.s.\ consistent and the assumptions of Proposition~\ref{Proposition:consitency} are fulfilled.
%The following result provides simple conditions so that $\Vert \hat s_1 - s_1 \Vert \to 0$ and $\Vert \hat s_2 - s_2 \Vert \to 0$ a.s. for the usual choices of target functions.
These results usually follow from Glivenko–Cantelli and the dominated convergence theorems.
%(under suitable integrability assumptions).
Corollary~\ref{Corollary-Consistency-s} considers 2DSD indices associated with the first, second and Lorenz dominance, but similar results can be derived for higher-degree stochastic orders.

%For $p\ge 1$, in what follows $\mathcal{L}^p$ is the usual space of random variables $X$ %on (the underlying probability space) $(\Omega,\mathcal{F}, \P)$
%such that $\E |X|^p<\infty$. In particular, $\mathcal{L}^1$ is the collection of integrable random variables and $\mathcal{L}^2$ is constituted by variables with finite second moment.

\begin{corollary}\label{Corollary-Consistency-s}
For the 2DSD indices associated with the usual stochastic order ($\mathcal{I}_1$ in \eqref{2DSD-1}), the Lorenz order ($\mathcal{I}_\ell$ in \eqref{2DSD-Lorenz}) and the second stochastic dominance  ($\mathcal{I}_2$ in \eqref{2DSD-2}), we have that:
  \begin{enumerate}[label={(\alph*)}, topsep=0mm, partopsep=0mm, itemsep=0mm, parsep=0mm, align=left, labelwidth=*, leftmargin=0 mm]
    \item\label{item:sc-1} If $\E |X_j| <\infty$, $j=1,2$, then $\hat\I_1 (X_1,X_2) \to \mathcal{I}_1 (X_1,X_2)$ and $\hat\I_\ell (X_1,X_2) \to \mathcal{I}_\ell (X_1,X_2)$  a.s.%, where $\mathcal{I}_1$ is the 2DSD index for the usual stochastic order in \eqref{2DSD-1} and $\mathcal{I}_\ell$ is the 2DSD index for the Lorenz order in \eqref{2DSD-Lorenz}.
    \item\label{item:sc-2} If $\E X_j^2 <\infty$, $j=1,2$, and $\E X_1=\E X_2$, we have that $\hat\I_2 (X_1,X_2) \to \mathcal{I}_2 (X_1,X_2)$ a.s.%, where $\mathcal{I}_2$ is the 2DSD index for the second stochastic dominance in \eqref{2DSD-2}.
    \item\label{item:sc-3}  If $X_j$ have compact support, $j=1,2$, then $\hat\I_2 (X_1,X_2) \to \mathcal{I}_2 (X_1,X_2)$ a.s.
  \end{enumerate}
\end{corollary}

\subsection{Asymptotic distribution of the 2DSD index}

The next theorem provides necessary conditions so that the (normalized) plug-in estimator of the 2DSD index, $\hat\I_s$ in \eqref{estimators-indices}, converges in distribution. This convergence
follows from the weak convergence of the process $\hat{s}_1-\hat{s}_2$ in the space $\mathrm{L}^1$ together with an extended version of the functional delta method; see \cite{Shapiro-1991}. %Only the convergence of  the underlying process is required to obtain the asymptotic distribution of the plug-in estimator of the index.
We use the arrow `$\rightsquigarrow$' to denote the weak convergence of probability measures in the underlying space.

%sense of Hoffmann-J{\o}rgensen (see \cite{van der Vaart-Wellner}) and the usual convergence in distribution for random vectors.

\begin{theorem} \label{Thm-AsympDistrIndex}
Assume that, as $n_1$, $n_2\to\infty$, it holds that
\begin{equation}\label{convergence-process}
%\sqrt{n_1n_2/(n_1+n_2)} (\hat{s}_1-\hat{s}_2 - ( s_1-s_2)) \rightsquigarrow \mathbb{S},\quad \text{in } \mathrm{L}^1.
r_{n_1,n_2}(\hat{s}_1-\hat{s}_2 - ( s_1-s_2)) \rightsquigarrow \mathbb{S},\quad \text{in the space } \mathrm{L}^1,
\end{equation}
for some stochastic process \(\mathbb{S}\) and some sequence $r_{n_1,n_2}\to \infty$.
Then, we have that
\begin{equation} \label{I0-normalized}
r_{n_1,n_2}   \left(\I (\hat s_1 ,\hat s_2 ) - \I ( s_1 , s_2 )\right)\rightsquigarrow \left( \int_{D_s} \mathbb{S} ,
  %\sqrt{\frac{n_1n_2}{n_1+n_2}}  \left(\I (\hat s_1 ,\hat s_2 ) - \I ( s_1 , s_2 )\right)\rightsquigarrow \left( \int_{D_s} \mathbb{S} ,
  \int_{\{s_1=s_2 \}} | \mathbb{S} |    +   \int_{\{s_1\ne s_2 \}}    \mathbb{S} \cdot \sgn(s_1-s_2)  \right),
\end{equation}
where $\sgn(\cdot)$ is the sign function.
\end{theorem}

In all the examples in this work the normalizing sequence in \eqref{convergence-process} and \eqref{I0-normalized}  is $r_{n_1,n_2} = \sqrt{{n_1n_2}/{(n_1+n_2)}}$. However, this rate might be different in other cases than those considered here.
%In the previous theorem, the convergence rate in \eqref{convergence-process} and \eqref{I0-normalized} is $\sqrt{{n_1n_2}/{(n_1+n_2)}}$. However, this rate might be different in other examples than those considered here. For simplicity we have taken this `standard' rate, but the same result holds for other possible normalizing sequences.
The following corollary provides necessary and sufficient conditions for the limit distribution in \eqref{I0-normalized} to be bivariate normal.

\begin{corollary}\label{Corollary-normal}
Under the condition of Theorem~\ref{Thm-AsympDistrIndex}, let us further assume that the limiting process $\mathbb{S}$ in \eqref{convergence-process} is Gaussian. The following three assertions are equivalent:
\begin{enumerate}[label={(\alph*)}, topsep=0mm, partopsep=0mm, itemsep=0mm, parsep=0mm, align=left, labelwidth=*, leftmargin=0 mm]
  \item\label{item:Corollary.normal.a}  The limit in \eqref{I0-normalized} has a bivariate normal distribution.

  \item\label{item:Corollary.normal.b} The process $\mathbb{S}\cdot 1_{\{s_1=s_2\}} = 0$ a.e.\ on $D_s$ almost surely.

   \item\label{item:Corollary.normal.c} The limit distribution in \eqref{I0-normalized} can be expressed as
 \begin{equation}\label{limit-distribution-2}
    I= \left( \int_{D_s} \mathbb{S} , \int_{D_s} \mathbb{S} \cdot \sgn(s_1-s_2) \right).
  \end{equation}
\end{enumerate}
In particular, if the set $\{s_1=s_2\} = \{ x\in D_s : s_1(x)=s_2(x) \}$ has zero Lebesgue measure, then either of the above statements is true.
\end{corollary}

%\begin{corollary}\label{Corollary-normal}
%Under the condition of Theorem~\ref{Thm-AsympDistrIndex}, let us further assume that the limiting process $\mathbb{S}$ in \eqref{convergence-process} is Gaussian. The following three assertions are equivalent:
%\begin{enumerate}[(a)]
%  \item\label{item:Corollary.normal.a} The set $\{s_1=s_2\} = \{ x\in D_s : s_1(x)=s_2(x) \}$ has zero Lebesgue measure.
%  \item\label{item:Corollary.normal.b} The limit distribution in \eqref{I0-normalized} can be expressed as
%  \begin{equation}\label{limit-distribution-2}
%    I= \left( \int_{D_s} \mathbb{S} , \int_{D_s} \mathbb{S} \cdot \sgn(s_1-s_2) \right).
    %I= \left( \int_{D_s}  \mathbb{S}(t)\,\d t, \,      \int_{D_s}  \mathbb{S}(t) \cdot \sgn(s_1(t)-s_2(t))  \,\d t   \right).
%  \end{equation}
  %where $\mathbb{S}$ is defined in \eqref{convergence-process}.
%  \item\label{item:Corollary.normal.c} The limit in \eqref{I0-normalized} has a bivariate normal distribution.
%\end{enumerate}
%\end{corollary}

%{blue}{Condition (\ref{item:Corollary.normal.a}) in Corollary \ref{Corollary-normal} appears several times in this work.
Observe that $\{s_1=s_2\}$ is the \textit{contact set} between $s_1$ and $s_2$. %, that is, the collection of the \textit{contact points} or \textit{crossing points} of the two target functions.
The fact that this set has measure zero means that the two target functions are essentially different.

To apply Theorem \ref{Thm-AsympDistrIndex} or the subsequent Corollary \ref{Corollary-normal}, we first need to check the convergence in \eqref{convergence-process}. This has to be established for each particular choice of $s$. In the following proposition we give conditions under which this weak convergence is fulfilled for the examples considered above. For the sake of clarity, we enumerate below all the conditions needed, although not all of them are required simultaneously in each example.

\begin{assumption}[\textbf{Sampling condition}]\label{Assumption-sizes}
  The sample sizes $n_1$ and $n_2$ satisfy that, as $n_1,n_2\to \infty$,  $n_1/(n_1+n_2)\to \lambda$, where $\lambda\in[0,1]$.
%\begin{equation}\label{sample-sizes}
%\frac{n_1}{n_1+n_2}\to \lambda,\quad \text{as }n_1,n_2\to \infty.
%\end{equation}
\end{assumption}

\begin{assumption}[\textbf{Integrability condition 1}] \label{Assumption-integrability-1}
For $j=1,2$, it holds that $\Lambda_{2,1}(X_j)<\infty$, where
  \begin{equation}\label{Delta-21}
  \Lambda_{2,1}(X_j)=\int_{0}^{\infty} \sqrt{\P(|X_j|>x)} \,\d x.
\end{equation}
\end{assumption}

\begin{assumption}[\textbf{Integrability condition 2}] \label{Assumption-integrability-2}
For $j=1,2$, it holds that $\Lambda_{4,2}(X_j)<\infty$, where
  \begin{equation}\label{Delta-42}
  \Lambda_{4,2}(X_j)=\int_{0}^{\infty} x \sqrt{\P(|X_j|>x)} \,\d x.
\end{equation}
\end{assumption}

\begin{assumption}[\textbf{Regularity condition}]\label{Assumption-regularity} For $j=1,2$, we have that $F_j(0)=0$ and $F_j$ has at most finitely many jumps and is continuously differentiable elsewhere with
strictly positive density.
\end{assumption}

Assumption \ref{Assumption-sizes} means that the sampling scheme is (weakly) balanced.
%The extreme values $\lambda=0,1$ correspond to the least interesting cases from a statistical point of view, as one of the populations is not asymptotically represented in the sample.
Assumption \ref{Assumption-integrability-1} amounts to saying that the variables $X_j$ belong to the Lorentz space $\mathcal{L}^{2,1}$ (see \citet[Section 1.4]{Grafakos}) and it is equivalent to the convergence of the empirical process in $\mathrm{L}^1$ (see \citet{del Barrio}).
Condition $\Lambda_{2,1}(X_j)<\infty$ is slightly stronger than $\E X_j^2<\infty$; it holds when $\E X_j^{2+\epsilon}<\infty$, for some $\epsilon>0$. Assumption \ref{Assumption-integrability-2} says that $X_j$ is in the Lorentz space $\mathcal{L}^{4,2}$, which is equivalent to the convergence of the integrated empirical process in $\mathrm{L}^1$; see \citet{Carcamo}. Again, this condition is a bit stronger than $\E X_j^4<\infty$.
%We observe that
These integrability conditions are optimal to obtain the convergence of the underlying processes. Other works on this subject assume that variables have compact support (see \cite{Huang-Kan-Tzeng-Wang-2021}), a much more demanding requirement. Finally, Assumption \ref{Assumption-regularity} is necessary to derive the convergence of the quantile process in $\mathrm{L}^1$ through the Hadamard differentiability of the inverse map plus the convergence of the empirical process in $\mathrm{L}^1$; see for example \cite{Sun-Beare-2021} and \cite{Kaji-2018}. This is used to derive the weak $\mathrm{L}^1$ convergence of the empirical Lorenz processes.

In the following proposition we show the convergence in $\mathrm{L}^1$ of the processes associated to the usual target functions. From now on, for $j=1,2$, $\B_{F_j}$ denote two independent \(F_j\)-Brownian bridges. %The proof is given in Section \ref{Section.Proofs}.

\begin{proposition}\label{Proposition.asymptitic}
If Assumption \ref{Assumption-sizes} holds, and for the target functions considered in Examples \ref{Example1stOrderSD}--\ref{ExampleLorenz}, we have the following:

\begin{enumerate}[label={(\alph*)}, topsep=0mm, partopsep=0mm, itemsep=0mm, parsep=0mm, align=left, labelwidth=*, leftmargin=0 mm]
    \item\label{item:Proposition.asymptotic.a} Under Assumption~\ref{Assumption-integrability-1}, it holds that, as \(n_1,n_2\to\infty\),
\begin{equation}\label{eq.conv.empirical.L1}
\sqrt{\frac{n_1n_2}{n_1+n_2}} (\hat{F}_1-\hat{F}_2 - ( F_1-F_2)) \rightsquigarrow  \sqrt{1-\lambda} \, \mathbb B_{F_1} - \sqrt{\lambda} \, \mathbb B_{F_2},\quad \text{in } \mathrm{L}^1.
\end{equation}

\item\label{item:Proposition.asymptotic.b}  Under Assumption~\ref{Assumption-integrability-2}, for positive  \(X_1\) and \(X_2\), it holds that, as \(n_1,n_2\to\infty\),
\[
\sqrt{\frac{n_1n_2}{n_1+n_2}} \left( \int_0^t (\hat{F}_1-\hat{F}_2) - \int_0^t( F_1-F_2)\right) \rightsquigarrow
\int_0^t ( \sqrt{1-\lambda} \, \mathbb B_{F_1}-  \sqrt{\lambda} \, \mathbb B_{F_2}),\quad \text{in } \mathrm{L}^1.
\]

\item\label{item:Proposition.asymptotic.c}  Under Assumption~\ref{Assumption-integrability-2}, for positive  \(X_1\) and \(X_2\), it holds that, as \(n_1,n_2\to\infty\),
\[
\sqrt{\frac{n_1n_2}{n_1+n_2}} \left( \int_t^\infty (\hat{F}_2-\hat{F}_1) - \int_t^\infty( F_2-F_1)\right) \rightsquigarrow
\int_t^\infty ( \sqrt{\lambda} \, \mathbb B_{F_2} - \sqrt{1-\lambda} \, \mathbb B_{F_1}),\quad \text{in } \mathrm{L}^1.
\]

\item\label{item:Proposition.asymptotic.d} Under Assumptions~\ref{Assumption-integrability-1} and \ref{Assumption-regularity}, for positive \(X_1\) and \(X_2\) with strictly positive means, it holds that, as \(n_1,n_2\to\infty\),
\[
\sqrt{\frac{n_1n_2}{n_1+n_2}} \left( \hat \ell_1 - \hat \ell_2 - (\ell_1-\ell_2) \right) \rightsquigarrow \sqrt{1-\lambda}\, \mathbb{L}_1-  \sqrt{\lambda}\,   \mathbb{L}_2,\quad \text{in } \mathrm{L}^1,
\]
where, for $j=1,2$, $\mathbb{L}_j$  are (independent) centered Gaussian processes with continuous trajectories a.s.\ that can be expressed as
\begin{equation}\label{Lorenz-convergence}
   \mathbb{L}_j(t)  =  \ell_j(t)\int_{0}^{1}  \ell^{\prime\prime}_j(x)\mathbb{B}_{F_j}(x)\, \d x  - \int_{0}^{t}  \ell^{\prime\prime}_j(x)\mathbb{B}_{F_j}(x)\, \d x,\quad t\in[0,1].
\end{equation}
\end{enumerate}
\end{proposition}

%The proof of part \emph{(\ref{item:Proposition.asymptotic.a})}

Under the assumptions of Lemma 2.1 in \cite{Sun-Beare-2021}, the result in Proposition \ref{Proposition.asymptitic} \textit{\ref{item:Proposition.asymptotic.d}} can be established in the space of continuous functions on $[0,1]$ with the supremum norm, which entails the convergence in $\mathrm{L}^1$.
Note that all the limiting process in Proposition \ref{Proposition.asymptitic} are centered Gaussian processes. Therefore, Corollary \ref{Corollary-normal} provides necessary and sufficient conditions so that the (normalized) estimator of the 2DSD index in the usual examples is asymptotically normal (with zero mean). % The following result considers the case of Examples \ref{Example1stOrderSD}--\ref{ExampleLorenz}.
Specifically, in Examples \ref{Example1stOrderSD}, \ref{Example2ndOrderSD} and \ref{ExampleStopLoss} consider $D_s$ as the union of the supports of the two distributions and assume that $\lambda\in(0,1)$ in Assumption \ref{Assumption-sizes}. In Example \ref{ExampleLorenz}, let $D_s=[0,1]$. In this situation, the limit in \eqref{I0-normalized} has a centered bivariate normal distribution if and only if the contact set $\{s_1=s_2\}$ has zero Lebesgue measure.

\subsection{Estimation of the minimum violation ratio \(\epsilon_0\)}

The literature on estimation of the MVR \(\epsilon_0\) in \eqref{MVR}
%, also called critical parameter,
is scarce. For the usual order, \cite{Bali-2009} and \cite{Levy-2012} outline an estimator based on the definition of ASD \eqref{ASD} and the plug-in principle.
\cite{Huang-Kan-Tzeng-Wang-2021} propose an estimator for \(\epsilon_0\) in a very specific context of insurance deductibles, for the second-degree SD and for compactly supported variables.

Here, we propose a consistent estimator of $\epsilon_0>0$ based on the empirical 2DSD index. Observe that, by equation \eqref{MVR.Characterization} in Proposition~\ref{Proposition.I.properties.1} \ref{item:Property.ASD}, we can express the MVR as
%\(\epsilon_0>0\) as
%\begin{equation*}%\label{MVR2}
%\epsilon_0 = \frac{1}{2}\left(   1- \frac{\int_{D_s}(s_1-s_2) }{\Vert s_1-s_2\Vert}   \right).
%\end{equation*}
%In particular, we have that
\(\epsilon_0 = g(\I(s_1,s_2))\) with \(g(x,y)=(1-x/y)/2\). Thus, a natural estimator of the MVR $\epsilon_0$ is
\begin{equation}\label{estimator-MVR}
\hat{\epsilon}_0  = g(\I(\hat s_1,\hat s_2))= \frac{1}{2} \left( 1- \frac{\int_{D_s} (\hat{s}_1-\hat{s}_2)}{\int_{D_s} |\hat{s}_1-\hat{s}_2|} \right).
%\frac{1}{2}\left(   1- \frac{\int_{D_s}(\hat{s}_1-\hat{s}_2) }{\Vert \hat{s}_1-\hat{s}_2\Vert}   \right).
\end{equation}
When \(\Vert s_1-s_2\Vert>0 \), then $g$ is differentiable at $\I(s_1, s_2)$, so the strong consistency  of \eqref{estimator-MVR} follows from Proposition~\ref{Proposition:consitency} and Corollary~\ref{Corollary-Consistency-s}. Moreover, whenever \(\epsilon_0>0\), the asymptotic distribution of (a standardized version of) $\hat{\epsilon}$ can be derived from Theorem~\ref{Thm-AsympDistrIndex} by application of the (first-order) delta method. Specifically, under the assumptions of Theorem~\ref{Thm-AsympDistrIndex} and when $\epsilon_0>0$, some simple computations show that
\begin{equation*}
  r_{n_1,n_2}(\hat\epsilon_0-\epsilon_0)\rightsquigarrow \frac{1}{2 \Vert s_1-s_2\Vert}  \left( (1-2\epsilon_0) \left( \int_{\{s_1=s_2 \}} |\mathbb{S}| + \int_{\{s_1\ne s_2 \}} \mathbb{S} \cdot \sgn(s_1-s_2) \right) - \int_{D_s} \mathbb{S} \right).
\end{equation*}
%\begin{equation*} r_{n_1,n_2}(\hat\epsilon_0-\epsilon_0)\rightsquigarrow \mathbb{E},
%\end{equation*}
%where
%\begin{equation*}
%\mathbb{E} = \frac{1}{2 \Vert s_1-s_2\Vert}  \left( (1-2\epsilon_0) \left( \int_{\{s_1=s_2 \}} |\mathbb{S}| + \int_{\{s_1\ne s_2 \}} \mathbb{S} \cdot \sgn(s_1-s_2) \right) - \int_{D_s} \mathbb{S} \right).
%\end{equation*}

%then \(\epsilon_0\) and \(\hat{\epsilon}_0\) are differentiable functions of the population and empirical 2DSD index, respectively, so the strong consistency of \eqref{estimator-MVR} follows from Proposition~\ref{Proposition:consitency} and Corollary~\ref{Corollary-Consistency-s}. Moreover, in this case, the asymptotic distribution of (a standardized version of) $\hat{\epsilon}^*$ can be derived from Theorem~\ref{Thm-AsympDistrIndex} together with the delta method.
%Details are omitted.

%where $\I_s(X_1,X_2)=(x,y)$.

%%%%%%%%%%%%%%%%%%%%%%%%%%%%%%%%%%
% Subsection Bootstrap consistency
%%%%%%%%%%%%%%%%%%%%%%%%%%%%%%%%%%
\subsection{Bootstrap consistency} \label{Subsect.BootConsist}

From the theoretical point of view, the previous consistency and asymptotic results ensure that the estimators \eqref{estimators-indices} and \eqref{estimator-MVR} are statistically reasonable and provide the precise rates of convergence. However, the corresponding limits are not distribution-free: they depend on the underlying (and usually unknown) populations $F_1$ and $F_2$. Moreover, even in the case where the limiting variable is bivariate normal (see Corollary \ref{Corollary-normal}), simulations show that convergence to the limit is relatively slow. Therefore, to make inferences on the 2DSD index \(\I_s\) in \eqref{2DSD-index} and the MVR \(\epsilon_0\) in \eqref{MVR}, we resort to a bootstrap scheme; see Section~\ref{Section.ASDtest}.

Here, we give general conditions that guarantee the strong consistency of the bootstrap estimators of the 2DSD index and MVR.
From Theorem \ref{Thm-AsympDistrIndex} and Corollary \ref{Corollary-normal}, we see that the asymptotic distribution of the 2DSD index changes radically depending on whether the contact set $\{s_1=s_2\}$ has measure zero or not.
This phenomenon has been observed previously. For tests of stochastic dominance as in \eqref{Dtest}, the limiting distribution of the test statistic in \cite{Linton-Song-Whang-2010} is discontinuous and hence not locally uniformly regular; see \cite{Bickel-1993}. For this reason we have divided this section into two parts depending on whether the contact set $\{s_1=s_2\}$ has zero or positive measure.

\subsubsection*{Case 1: The contact set has zero measure}

The set of crossing points between two target functions usually consists of a finite (or empty) collection of points.
%For example, if $s$ is the Lorenz curve in \eqref{Lorenz-curve} (see Example \ref{ExampleLorenz}), any small variation in the distribution of wealth (even within the same country) %(e.g., any transfer as in the Pigou-Dalton principle)
%generally entails a change in the mean and this yields a completely different Lorenz curve. If $s_1$ and $s_2$ are generated by continuous random mechanisms, the probability that these functions overlap in a non-empty open interval is normally negligible.
Therefore, in the usual practical situations, the contact set has zero Lebesgue measure and the results of this part are particulary useful.
For each $j=1,2$, the bootstrap sample $\{X_{j,i}^*\}_{i=1}^{n_j}$ is obtained by redrawing independently from the empirical distribution $\hat F_j$ in \eqref{empirical.distribution.function}. The estimator of \(s_j\) derived from this bootstrap sample is denoted by \(\hat s_j^*\).
%the bootstrap process (associated with $s_j$) is
%\begin{equation}\label{Bootstrap-S}
%\hat{\mathbb S}_j^* = \sqrt{n_j}(\hat s_j^*-\hat s_j), \quad j=1,2.
%\end{equation}
%The following result provides necessary and sufficient conditions for a standard bootstrap scheme to work.

\begin{theorem}\label{Theorem.boostrap.general}
Let the assumptions in Theorem~\ref{Thm-AsympDistrIndex} hold, where \(\mathbb{S}\) is a Gaussian process with support $D_{\mathbb{S}}$.
Assume also that the bootstrap process
\begin{equation}\label{convergence-processBOOT}
%\sqrt{n_1n_2/(n_1+n_2)} (\hat{s}_1^*-\hat{s}_2^* - ( \hat s_1-\hat s_2))
r_{n_1,n_2} (\hat{s}_1^*-\hat{s}_2^* - ( \hat s_1-\hat s_2))
\end{equation}
is a.s.\ consistent in $\mathrm{L}^1$, that is, conditionally to $\{X_{j,i}\}_{i=1}^{n_j}$ ($j=1,2$), it converges weakly in \(\mathrm{L}^1\) to \(\mathbb{S}\) a.s.
Then, the following two assertions are equivalent:
\begin{enumerate}[label={(\alph*)}, topsep=0mm, partopsep=0mm, itemsep=0mm, parsep=0mm, align=left, labelwidth=*, leftmargin=0 mm]
\item\label{item:Theorem.boostrap.general.a} The set $\{s_1=s_2\}\cap D_{\mathbb{S}} = \{ x\in D_s\cap D_{\mathbb{S}}  : s_1(x)=s_2(x) \}$ has zero Lebesgue measure.
\item The bootstrap estimator \(\I (\hat s_1^* ,\hat s_2^* )\) of the 2DSD index is a.s.\ consistent.
\end{enumerate}
In particular, if any of the previous two conditions holds, then the bootstrap estimator $\hat\epsilon_0^* = g(\I(\hat s_1^*,\hat s_2^*))$ of the MVR $\epsilon_0>0$ is also a.s.\ consistent.
\end{theorem}

The proof of Theorem \ref{Theorem.boostrap.general} follows from the full Hadamard differentiability of the involved mappings and \citet[Theorem 3.1]{Fang-Santos}. As in the previous results, the strong $\mathrm{L}^1$-consistency of the bootstrap process in (\ref{convergence-processBOOT}) has to be established for the selected \(s\). %We we adapt (to the space $\mathrm{L}^1$) the ideas of the consistency of the empirical bootstrap process for the supremum norm; see \citet[Section 3.6]{van der Vaart-Wellner}.
The following proposition shows the strong consistency (in $\mathrm{L}^1$) of the bootstrap processes underlying the estimated indices in Examples \ref{Example1stOrderSD}--\ref{ExampleLorenz}.
The proof of this result can be derived from \cite[Theorem 2.4 and Remark 2.5]{Gine-Zinn}, where general conditions for the consistency of the bootstrap process in a Banach space are established. However, in the online Appendix we include a simpler alternative proof of this result for the considered examples. In particular, together with Theorem \ref{Theorem.boostrap.general}, we conclude that the corresponding bootstrapped 2DSD indices and MVR's are consistent a.s. The precise definitions of the involved processes can be found in the online Appendix.

\begin{proposition} \label{Proposition.ConsistBOOtEmpLorProc}
Under the assumptions in Proposition \ref{Proposition.asymptitic}, we have that the empirical, integrated empirical and empirical Lorenz bootstrap processes are $\mathrm{L}^1$-consistent a.s.
\end{proposition}

\subsubsection*{Case 2: The contact set has positive measure}%. Background and motivation} \color{black}

Theorem \ref{Theorem.boostrap.general} shows that the standard bootstrap is no longer valid to estimate the asymptotic distribution of the 2DSD index whenever $\{s_1=s_2\}$ has positive measure. In such a case, the $\mathrm{L}^1$-norm (second component of the 2DSD index in \eqref{2DSD-index}) is non-differentiable at $s_1-s_2$; see Lemma 1 in the online Appendix or \citet[Lemma 4]{Carcamo}.
In this situation, \cite[Section 3.3]{Fang-Santos} provide different consistent modifications of the bootstrap scheme. The idea is that, although the $\mathrm{L}^1$-norm is not fully differentiable, it is \textit{Hadamard directionally differentiable} and the directional derivative can be appropriately estimated to obtain a consistent alternative.

As suggested in \cite[p.\ 391]{Fang-Santos}, it is convenient to employ an estimator of the directional derivative based on its analytical expression to construct a consistent resampling scheme. In the following result we use the explicit expression of the derivative of the $\mathrm{L}^1$-norm (given in the second component of the limit in \eqref{I0-normalized}) together with the ideas in \cite{Linton-Song-Whang-2010} about \textit{enlargements} of the contact set. As a result, we obtain a consistent resampling method for statistics defined through the $\mathrm{L}^1$-norm. As this result might be of interest in other situations, we state the next proposition in general terms: $\theta_0\in \mathrm{L}^1$, $\hat{\theta}_n$ is an estimator of $\theta_0$, and  $\hat{\theta}_n^*$ its bootstrapped version.

\begin{theorem}\label{Theorem.Modified.bootstrap}
Assume that the following two conditions hold:
\begin{enumerate}[label={(\roman*)}, topsep=0mm, partopsep=0mm, itemsep=0mm, parsep=0mm, align=left, labelwidth=*, leftmargin=0 mm]
  \item There exists a sequence $r_n\uparrow\infty$ such that $r_n (\hat{\theta}_n - \theta_0)\rightsquigarrow \mathbb{G}_0$ in $\mathrm{L}^1$, where $\mathbb{G}_0$ is a tight random element in $\mathrm{L}^1$.
  \item The bootstrap process $r_n (\hat{\theta}_n^*-\hat{\theta}_n)$ is a.s.\ consistent in $\mathrm{L}^1$.
\end{enumerate}
Then, the distribution of $\hat{\delta}'_n(r_n(\hat{\theta}_n^*-\hat{\theta}_n))$ conditioned to the observed sample is a consistent estimator for the asymptotic distribution of $r_n(\Vert \hat{\theta}_n\Vert - \Vert \theta_0\Vert)$, where
\begin{equation}\label{estimated.derivativa}
\hat{\delta}'_n(h)  =  \int_{\{ |\hat{\theta}_n|\le a_n \}} |h| + \int_{\{ |\hat{\theta}_n|>a_n  \}} h \cdot \sgn(\hat{\theta}_n),\quad h\in\mathrm{L}^1,
\end{equation}
and $(a_n)$ is any sequence of real numbers such that $a_n\downarrow 0$ and $a_n r_n\uparrow\infty$.
\end{theorem}

The quantity $a_n$ in Theorem \ref{Theorem.Modified.bootstrap} plays the role of a \textit{step size} or \textit{tuning parameter} in the approximation of the derivative.
Usually, $r_n = \sqrt{n}$, and in this case typical choices of $a_n$ are $a_n = c/\log n$ or $c/\log\log n$, with $c>0$ a constant.
%In problems with two populations, as those considered in this work, we can apply Theorem \ref{Theorem.Modified.bootstrap} to $\theta_0=s_1-s_2$ and $r_n=\sqrt{n_1n_2/(n_1+n_2)}$ to obtain a consistent resampling method to approximate the asymptotic distribution.% of \(\hat \I\).
%the estimated 2DSD index in \eqref{estimators-indices}.

  Theorem~\ref{Theorem.Modified.bootstrap} provides a consistent resampling scheme for all the tests in \eqref{Test.MVR}. For instance, following the notation in \citet[Sec. 3.4.2]{Fang-Santos}, from \eqref{MVR.Characterization}, the test \eqref{Test.MVR}(a) can be restated as
\begin{equation} \label{ASDTest_ContactSet}
H_0:\phi(\theta_0)\leq 0 \quad\text{versus}\quad  H_1:\phi(\theta_0)>0,
\end{equation}
where \(\theta_0=s_1-s_2\) and
\begin{equation}\label{phi}
\phi(h) = \int_{D_s} h - (1-2\epsilon) \|h\|,\quad h\in\mathrm{L}^1.
\end{equation}
As the first summand of $\phi$ in \eqref{phi} is linear, we can use Theorem \ref{Theorem.Modified.bootstrap} to obtain a suitable bootstrap estimator of $\phi(\theta_0)$.
Specifically, for \(\hat\theta_n=\hat s_1-\hat s_2\) and \(r_n=\sqrt{n_1n_2/(n_1+n_2)}\), we consider
\begin{equation}\label{modified-bootstrap}
\hat\phi_n'(h) = \int_{D_s} h - (1-2\epsilon) \hat{\delta}'_n(h)= \int_{D_s} h - (1-2\epsilon) \left( \int_{\{ |\hat{\theta}_n|\le a_n \}} |h| + \int_{\{ |\hat{\theta}_n|>a_n  \}} h \cdot \sgn(\hat{\theta}_n) \right),
\end{equation}
with $(a_n)$ being any sequence of real numbers such that $a_n\downarrow 0$ and $a_n r_n\uparrow\infty$. By Theorem \ref{Theorem.Modified.bootstrap}, the distribution of \(\hat\phi_n'(r_n(\hat\theta_n^*-\hat\theta_n))\), conditional on the sample, is a consistent estimator of \(r_n(\phi(\hat\theta_n)-\phi(\theta_0))\).

%By Lemma 1 in the online appendix, the Hadamard directional derivative of \(\phi\) at \(\theta\) is
%\begin{equation} \label{DirectionalDerivative}
%\phi_\theta'(h) = \int_{D_s} h - (1-2\epsilon) \left( \int_{\{\theta=0\}} |h| + \int_{\{\theta\neq 0\}} h \, \sgn(\theta) \right).
%\end{equation}

\section{Testing almost stochastic dominance} \label{Section.ASDtest}

After obtaining the asymptotic distribution of the empirical index (Theorem \ref{Thm-AsympDistrIndex} and Corollary \ref{Corollary-normal}), it seems natural to use these results to derive a rejection region for the ASD tests in \eqref{Test.MVR}. However, simulations show that the distribution of the standardized estimated 2DSD index is only approximately Gaussian for very large sample sizes. Then, for moderate sample sizes, asymptotic distributions could differ significantly from the corresponding finite sample distributions. Theorems \ref{Theorem.boostrap.general} and \ref{Theorem.Modified.bootstrap} together with Proposition \ref{Proposition.ConsistBOOtEmpLorProc} guarantee that suitable bootstrap rejection regions for the tests in \eqref{Test.MVR} are consistent. We have to distinguish between the two cases indicated in Section~\ref{Subsect.BootConsist}.

\subsubsection*{Case 1: The contact set has zero measure}

Assuming that the set \(\{s_1=s_2\}\) has zero measure, we propose a standard bootstrap algorithm to construct rejection regions for the tests
\eqref{Test.MVR}, at a significance level \(\alpha\), where \(\epsilon\) is a pre-specified violation ratio of interest:

\textbf{Step 1.} Extract \(B\) bootstrap samples from each of the original samples.
\begin{equation} \label{flowchartASDtest}
    \begin{array}{cccc}
\mbox{Original samples} & & \mbox{Bootstrap samples} \\
\begin{array}{l}
x_{11},\ldots,x_{1n_1} \\ x_{21},\ldots,x_{2n_2}
\end{array}
&
\begin{array}{c}
\longrightarrow \\ \longrightarrow
\end{array}
&
\begin{array}{l}
x_{11}^{*b},\ldots,x_{1n_1}^{*b} \\ x_{21}^{*b},\ldots,x_{2n_2}^{*b}
\end{array}
&
b=1,\ldots,B.
\end{array}
\end{equation}
\textbf{Step 2.} For each \(b=1,\ldots,B\) and the corresponding pair of bootstrap samples \eqref{flowchartASDtest}, compute the bootstrapped empirical inequality index \(\hat{\mathcal I}^{*b}\) and MVR \(\hat\epsilon_0^{*b}\).
%as in \eqref{estimators-indices} and \eqref{estimator-MVR}.

\textbf{Step 3.} Denote by \(\hat\epsilon_0^{*(\alpha)}\) the \(\alpha\)-quantile of the bootstrapped MVR's \(\hat\epsilon_0^{*1},\ldots,\hat\epsilon_0^{*B}\). We reject the null hypothesis in the test given in 
\begin{enumerate}[topsep=0mm, partopsep=0mm, itemsep=0mm, parsep=0mm, labelindent=0mm, align=left, labelwidth=*, leftmargin=0 mm]
\item[(\ref{Test.MVR}a)] %when at least a proportion \(1-\alpha\) of the bootstrapped MVR's are less than or equal to \(\epsilon\), that is,
    when \(\epsilon\geq\hat\epsilon_0^{*(1-\alpha)}\).
    %. In the absence of a reference \(\epsilon\) for which the practitioner wants to carry out the ASD test, the quantile \( \hat\epsilon_0^{*(1-\alpha)} \)
    This quantile provides the infimum of the \(\epsilon\) for which \(H_0\)
    %in the test (\ref{Test.MVR}a)
    is rejected at level \(\alpha\) and the endpoint of a one-sided confidence interval for \(\epsilon_0\) at level \(1-\alpha\).
\item[(\ref{Test.MVR}b)] when %at least a proportion \(1-\alpha\) of the bootstrapped MVR's are greater than or equal to \(\epsilon\), i.e.,
    \(\epsilon \leq \hat\epsilon_0^{*(\alpha)} \). This quantile is the left endpoint of the other one-sided confidence interval for \(\epsilon_0\) at level \(1-\alpha\). \color{black}
\item[(\ref{Test.MVR}c)] when \(\epsilon\) is not contained in the confidence interval for \(\epsilon_0\) given by \( [\hat\epsilon_0^{*(\alpha/2)},\hat\epsilon_0^{*(1-\alpha/2)}] \).
\end{enumerate}

\subsubsection*{Case 2: The contact set has positive measure}

In the general case when \(\{s_1=s_2\}\) might not have zero measure, we can use the modified bootstrap scheme obtain in Section~\ref{Subsect.BootConsist} based on the estimator
of the directional derivative in \eqref{modified-bootstrap}. This implementation is slightly more complicated than that of Case 1, because the testing procedure in Case 2 depends on fixing in advance the value of \(\epsilon\).
For example,  we reject the null hypothesis in  \eqref{ASDTest_ContactSet} (which is equivalent to (\ref{Test.MVR}a)) when the test statistic \(r_n\phi(\hat\theta_n)\) exceeds \(c_{1-\alpha}\), the \((1-\alpha)\)-quantile of the corresponding asymptotic distribution when \(\phi(\theta_0)=0\). %Next, we describe a resampling procedure to approximate this quantile.
%\textcolor{red}{(No he comprobado que se satisface la Hip\'{o}tesis 4 de \cite{Fang-Santos}.)}
Then, the critical value \(c_{1-\alpha}\) can be estimated by \(\hat c_{1-\alpha}\), the \((1-\alpha)\)-quantile of \(\hat\phi_n'(r_n(\hat\theta_n^*-\hat\theta_n))\), conditional on the sample, where $\hat\phi_n'$ is in \eqref{modified-bootstrap}. %This latter quantile can be approximated by resampling \(\hat\theta_n^*\).

We finally observe that the proof of Theorems~\ref{Theorem.boostrap.general} and \ref{Theorem.Modified.bootstrap} (see the online appendix) ensure that \cite[Theorem~3.3]{Fang-Santos} can be applied and this implies the pointwise size control of the testing procedures described in this section.

%%%%%%%%%%%%%%%%%%%%%%%%%%%%%%%%%%%%%%%%%%%%%%%%%%%%%%%%%%%%%%%%%%%%%%%%%%%%%%%%%%%
% SECTION: SIMULATIONS
%%%%%%%%%%%%%%%%%%%%%%%%%%%%%%%%%%%%%%%%%%%%%%%%%%%%%%%%%%%%%%%%%%%%%%%%%%%%%%%%%%%

\subsection*{A simulation study} \label{Section.Simulations}

%The above bootstrap rejection scheme has a good practical performance both in simulations and in the analysis of real data.
We present the results of a Monte Carlo study for the test (\ref{Test.MVR}a).
%Segundo p\'{a}rrafo de la pg 4743 de Huang et al 2021) ``{\em whether the acceptable violation ratio would have a value resembling that from the laboratory-experimental results}''}
%Section~\ref{Section.RealData} are dedicated to demonstrate the good practical performance of the bootstrap rejection scheme for the almost dominance test in (\ref{ASDtest-s}).
%Here, we summarize the results of a Monte Carlo study.
In Table~\ref{Table.SimulPops} we indicate the four scenarios considered, along with the MVR \(\epsilon_0\) and the stochastic orders with respect to which the test is carried out (graphical representations of the distribution functions or the Lorenz curves appear in the online appendix).
%First, we carry out this ASD test with respect to the usual order, sampling from lognormal distributions under Scenarios 1 and 2 in Table~\ref{Table.SimulPops}; see the cumulative distribution functions (cdf) in Supplementary Figure 1.
In Scenario 1, the distribution functions \(F_1\) and \(F_2\) intersect only once in the right tail
%(the crossing point is marked by a dashed vertical line)
and fulfill \(X_2 \preceq_{s}^\epsilon X_1\) for \(\epsilon\geq\epsilon_0\).
In Scenario 2, \(F_1\) and \(F_2\) cross once too, but near the origin, with a smaller MVR.
We also perform the ASD test with respect to the Lorenz order (Scenario 3), simulating samples from two generalized beta distributions of the second kind (GB2).
% because this distribution  is appropriate in the analysis of income and other positively skewed distributions; see \cite{Chotikapanich-et-al-2018}.
The scale parameter of the GB2 distribution is
%\(b_j = B(p_j,q_j)/B(p_j+1/a_j,q_j+1/a_j)\)
chosen so that \(\E(X_j)=1\), for \(j=1,2\).
%This last example corresponds to Scenario 3 in Table~\ref{Table.SimulPops}.
Finally, Scenario 4 is designed to illustrate Case 2, as the distribution functions \(F_1\) and \(F_2\) coincide over the interval $[0,1]$; see the specific expression of \(F_2\) in the appendix.

\begin{table} 
\begin{center}
    \begin{tabular}{cl|llcl}
    \mbox{} & \multicolumn{1}{c}{\mbox{}} & \multicolumn{1}{c}{Population 1} & \multicolumn{1}{c}{Population 2} &
    MVR \(\epsilon_0\) & Order \\ \hline
    \parbox[t]{2mm}{\multirow{4}{*}{\rotatebox[origin=c]{90}{Scenario}}} & 1 & LN(\(\mu_1=2\),\(\sigma_1=1\)) & LN(\(\mu_2=1\),\(\sigma_2=1.5\)) & 0.127290 & Usual \\
    & 2 & LN(\(\mu_1=1\),\(\sigma_1=1\)) & LN(\(\mu_2=1\),\(\sigma_2=1.5\)) & 0.036160 & Usual \\
    & 3 & GB2(\(a_1 = 2\), \(p_1 = 0.8\), \(q_1 = 1.5\)) & GB2(\(a_2 = 9\), \(p_2 = 0.1\), \(q_2 = 7\)) & 0.063151 & Lorenz \\
    & 4 & LN(\(\mu_1=1\),\(\sigma_1=1\)) & Composite & 0.081216 & Usual \\ \hline
    \end{tabular}
\end{center}
\caption{Models for the Monte Carlo study. N stands for normal, LN for lognormal and GB2 for generalized beta of the second kind. Population 2 in Scenario 4 is a composite distribution of a LN(1,1), a N(0,0.5) and a LN(1,1.1) (see the online appendix for the detailed expression).}
\label{Table.SimulPops}
\end{table}

\begin{figure}
    \centering
    \begin{tabular}{cc}
    \includegraphics[width=0.48\textwidth]{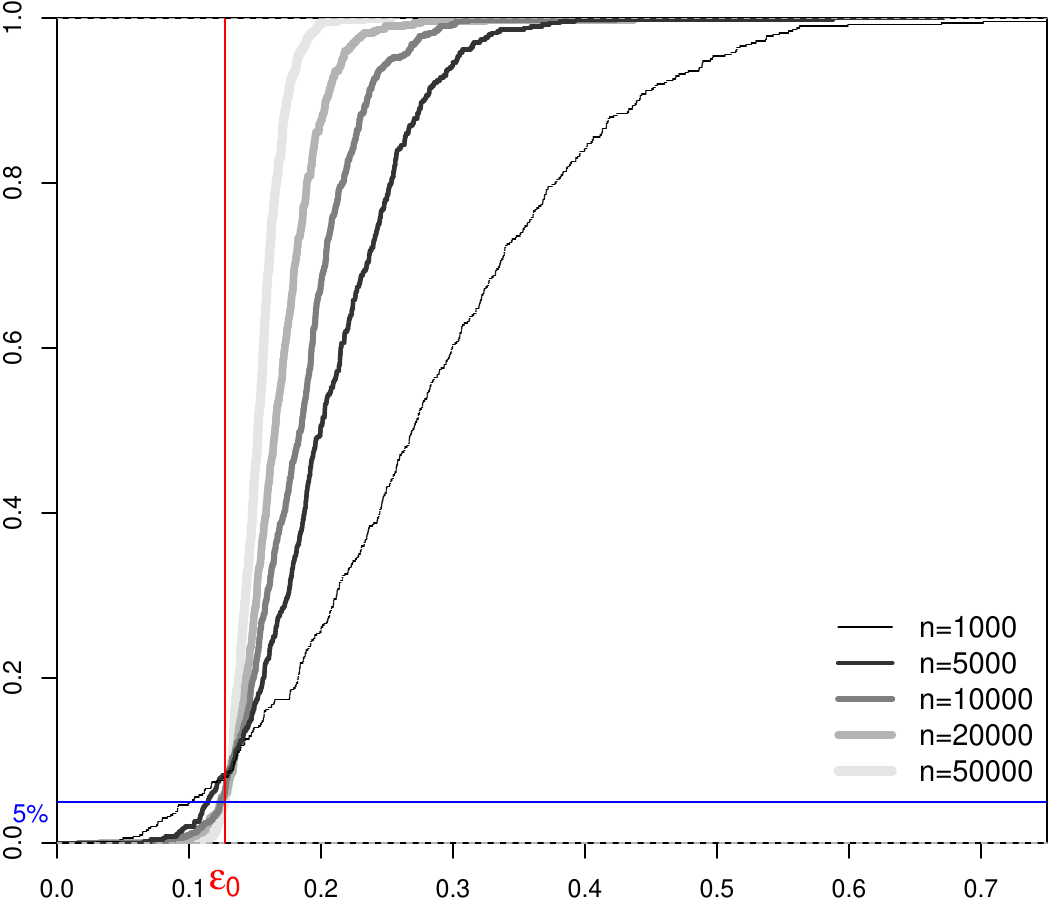} & \includegraphics[width=0.48\textwidth]{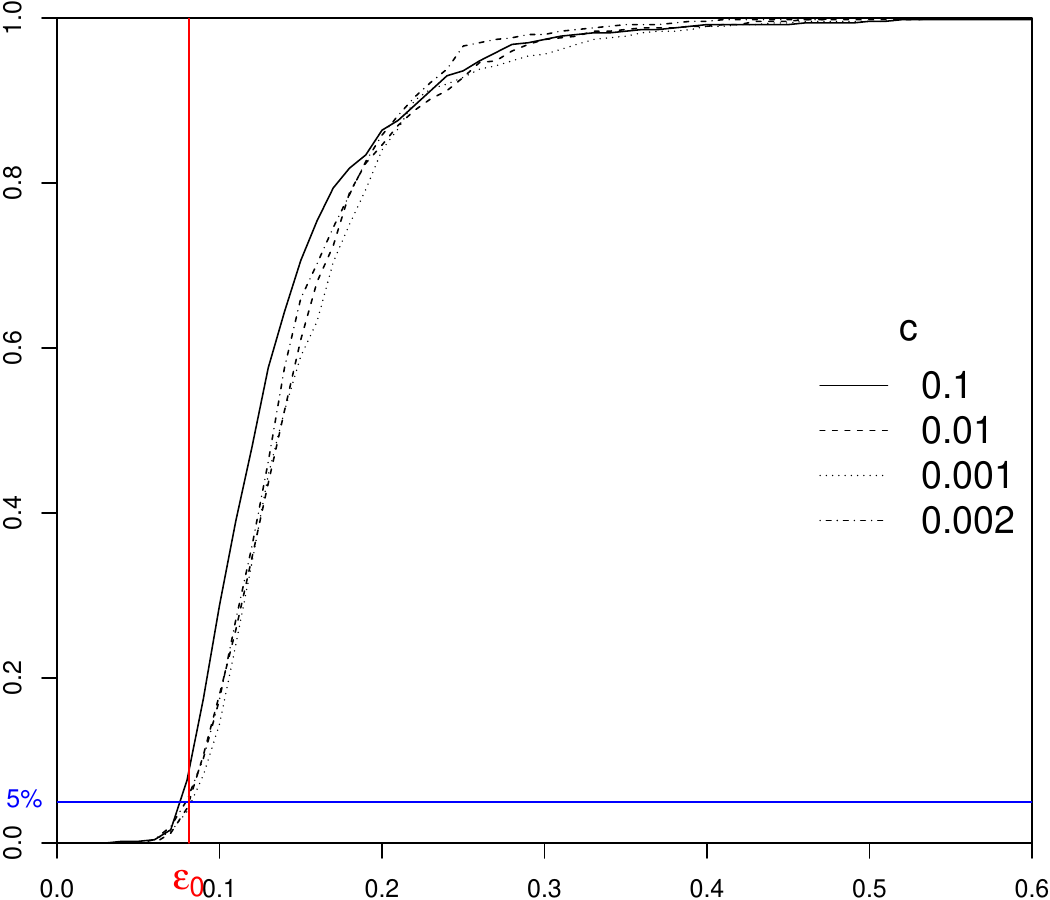} \\ (a) Scenario 1. & (b) Scenario 4 \\
    \end{tabular}
    %\includegraphics[width=0.6\textwidth]{SimModel2EpsQuantECDF.pdf} \\ (b) Scenario 2. \\
    %\caption{Ecdf's for the 500 Monte Carlo values of \( \hat\epsilon_0^{*(0.95)} \)  .}
    %\label{fig:SimulCase2EpsQuant}
%\end{figure}
%\begin{figure}
%    \centering
    %\includegraphics[width=0.6\textwidth]{SimGB2Model1EpsQuantECDF.pdf} \\ (c) Scenario 3.
    \caption{Power curve as a function of \(\epsilon\) for the test \(H_0:\epsilon_0\geq \epsilon\).}
    \label{fig:SimulCase3EpsQuant}
\end{figure}

The technical details of the simulation study are as follows. The significance level is \( \alpha=0.05 \). In every scenario we simulate \(N=500\) Monte Carlo samples (of the same size \(n_1=n_2=n\)) from each of the two models (populations 1 and 2). To examine the effect of the sample size, in Case 1 (resp. Case 2) we selected \(n\) as 1000, 5000, 10000, 20000 and 50000 (resp., 1000, 5000 and 10000). For each Monte Carlo run we drew \(B=2000\) bootstrap samples and computed \( \hat\epsilon_0^{*(0.95)} \) in Case 1.
With respect to Case 2, the sequence \(a_n\) used in \eqref{modified-bootstrap} is \(a_n = c \, \log(n_1n_2/(n_1+n_2))/r_n\) (similar to the choice in \cite{Linton-Song-Whang-2010}), with \(r_n=\sqrt{n_1n_2/(n_1+n_2)}\). In Case 2 the power of the testing procedure was evaluated for a grid of values of \(\epsilon\) in (0,1) and for different values of \(c\), namely 0.1, 0.01, 0.002 and 0.001.
Sampling from the composite distribution in Scenario 4 was carried out with the R library \texttt{mistr} (\cite{Sablica-Hornik}).
%This results in \(N\) simulated values of \( \hat\epsilon_0^{*(0.95)} \), whose 5\% quantile is reported in Table~\ref{Table.SimulResults} for sample sizes \(n\) = 1000, 5000, 10000, 20000 and 50000.
For Scenarios 1 and 4, Figure~\ref{fig:SimulCase3EpsQuant} displays, for the values of \(\epsilon\) in the horizontal axis, the power of the test~(\ref{Test.MVR}a), in Scenario 1 (Case 1) for the different values of the sample size \(n\) and in Scenario 4 (Case 2) for \(n=5000\) and the values of \(c\). The power curves corresponding to Scenarios 2 and 3 are analogous to those of Scenario 1 and appear in the online appendix, as well as the power curves for Scenario 4 when \(n=1000\) and \(n=10000\).

\color{black}

The simulations confirm the consistency of the bootstrap testing procedure as previously observed:
the asymptotic rejection probability is zero in the interior of the null and the asymptotic power is equal to 1 under the alternative hypothesis. Figure~\ref{fig:SimulCase3EpsQuant}(a) shows that, as the sample size \(n\) increases, the power approaches 1 when the alternative hypothesis is true and remains bounded by 5\% when the null is true. In the power functions for Scenarios 2 and 3 (online appendix), the significance level is very close to or below the 5\% nominal level for all values of the sample size \(n\).
Scenario 2 (resp., 3) corresponds to a crossing point of the distribution (resp., Lorenz) functions near the origin.
Since the crossing point in these two scenarios lies in a high density zone, there is more sample information available to better estimate that point and the MVR \(\epsilon_0\) when strict domination fails.
This is the opposite of Scenario 1, in which the two distribution functions \(F_1\) and \(F_2\) cross in a low density zone, and the estimated MVR depends on the integral of the difference \(F_1-F_2\) from the crossing point to infinity. Hence, it takes a larger \(n\) for the ecdf of Figure~\ref{fig:SimulCase3EpsQuant} (a) to approach the 5\% significance level under \(H_0\).
Taking \(B>2000\) bootstrap samples does not significantly improve the simulation results.
Regarding Scenario 4 (where the contact set has positive measure), we see that the bootstrap ASD testing procedure devised for Case 2 has a good performance and it is fairly robust with respect to the value of \(c\). Observe that a ``tightly'' estimated contact set (\(c=0.01\) for \(n=1000\), \(c=0.01\), 0.002 or 0.002 for \(n=5000\) and \(c=0.002\) for \(n=10000\)) leads to the power function adjusting better to $0.05$ in \(\epsilon=\epsilon_0\) and to 1 for \(\epsilon>\epsilon_0\).

%\begin{table}[]
%    \centering
%    \begin{tabular}{lrrrrr}
%           & \multicolumn{5}{c}{5\%-quantile of simulated \(\epsilon^*(0.05)\)'s} \\ \hline
%          \multicolumn{1}{r}{\(n=\)} & 1000 & 5000 & 10000 & 20000 & 50000 \\ \hline
%    Case 1 & 0.102445 & 0.126005 & 0.119928 & 0.120930 & 0.127017 \\
%    Case 2 & 0.037367 & 0.036031 & 0.036614 & 0.036280 & 0.036282 \\
%    Case 3 & 0.067325 & 0.064483 & 0.063646 & 0.063501 & 0.063804 \\ \hline
%    \end{tabular}
%    \caption{Results of the Monte Carlo study.}
%    \label{Table.SimulResults}
%\end{table}

%\begin{table}[]
%    \centering
%    \begin{tabular}{lrrrrr}
%           & \multicolumn{5}{c}{Significance level} \\ \hline
%          \multicolumn{1}{r}{\(n=\)} & 1000 & 5000 & 10000 & 20000 & 50000 \\ \hline
%    Case 1 & 0.080 & 0.068 & 0.074 & 0.070 & 0.052 \\
%    Case 2 & 0.038 & 0.052 & 0.030 & 0.042 & 0.048 \\
%    Case 3 & 0.036 & 0.044 & 0.046 & 0.050 & 0.040 \\ \hline
%    \end{tabular}
%    \caption{Results of the Monte Carlo study: empirical significance level.}
%    \label{Table.SimulResultsSignificance}
%\end{table}

%%%%%%%%%%%%%%%%%%%%%%%%%%%%%%%%%%%%%%%%%%%%%%%%%%%%%%%%%%%%%%%%%%%%%%%%%%%%%%%%%%%
% SECTION: TEST OF ALMOST LORENZ DOMINANCE IN REAL INCOME DATASETS
%%%%%%%%%%%%%%%%%%%%%%%%%%%%%%%%%%%%%%%%%%%%%%%%%%%%%%%%%%%%%%%%%%%%%%%%%%%%%%%%%%%

\section{Application to real data sets} \label{Section.RealData}

Here the bootstrap ASD testing procedures introduced in Section~\ref{Section.ASDtest} are illustrated with real data.  We consider two stochastic orders, the usual one (see Example~\ref{Example1stOrderSD}) and the Lorenz order (Example~\ref{ExampleLorenz}).
The results of this section, together with the simulations of Section~\ref{Section.ASDtest}, support the use of this general estimating and testing procedures for the minimum violation ratio \(\epsilon_0\) in practical, real data settings.

\subsection{The gender wage gap in Spain: first-order almost dominance} \label{Section.WageGAP}

The \textit{gender pay gap}, the difference in wages between women and men,
%; see \cite{Blau-Kahn-2017}.
has been analysed in various ways, but it is generally accepted that females systematically earn less than males.
It is natural to compare the salaries between both groups via their empirical distribution functions.
For instance, \cite{Maasoumi-Wang-2019} carry out first---and second---order SD tests as \eqref{Dtest} of female vs. male wage distributions in United States for all the years in the period 1976--2013. In 12 of these years the authors were unable to establish any of the two types of dominance and they suggest that, in this case, the claim that women are worse off than men is not robust.
However, a graphical representation of the pairs (women and men) of empirical distribution functions for some years gives insight into the relative position of these functions when strict dominance cannot be statistically accepted: there is a crossing of the distribution functions at the left tail (utmost lowest wages), while the cdf of male wages generally lies under the female one elsewhere.

\begin{figure}
    \centering
    \begin{tabular}{cc}
    \includegraphics[width=0.48\textwidth]{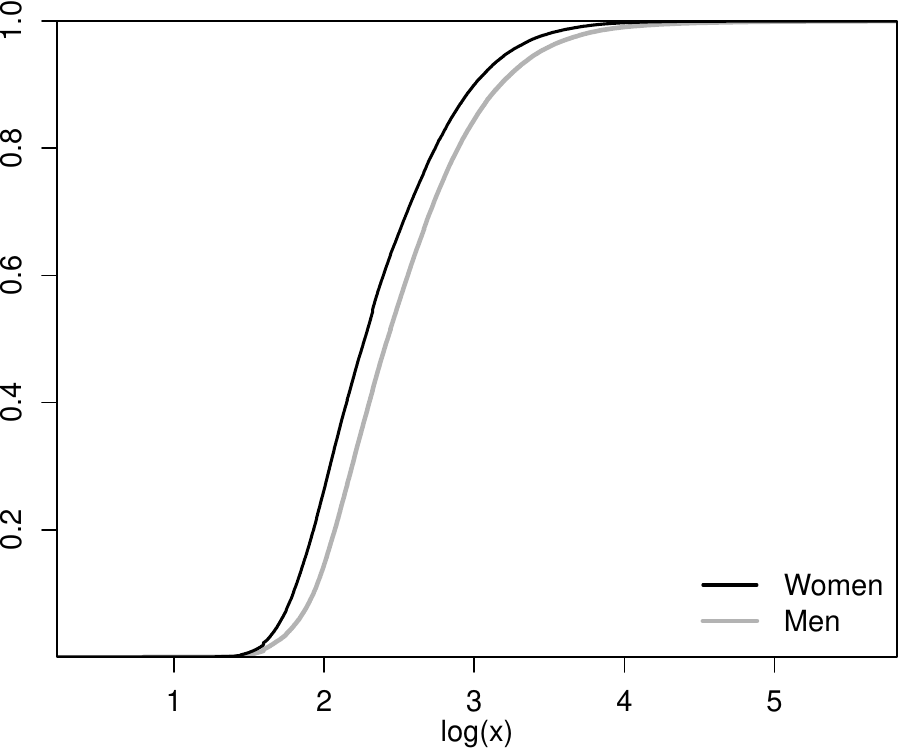} &
    \includegraphics[width=0.48\textwidth]{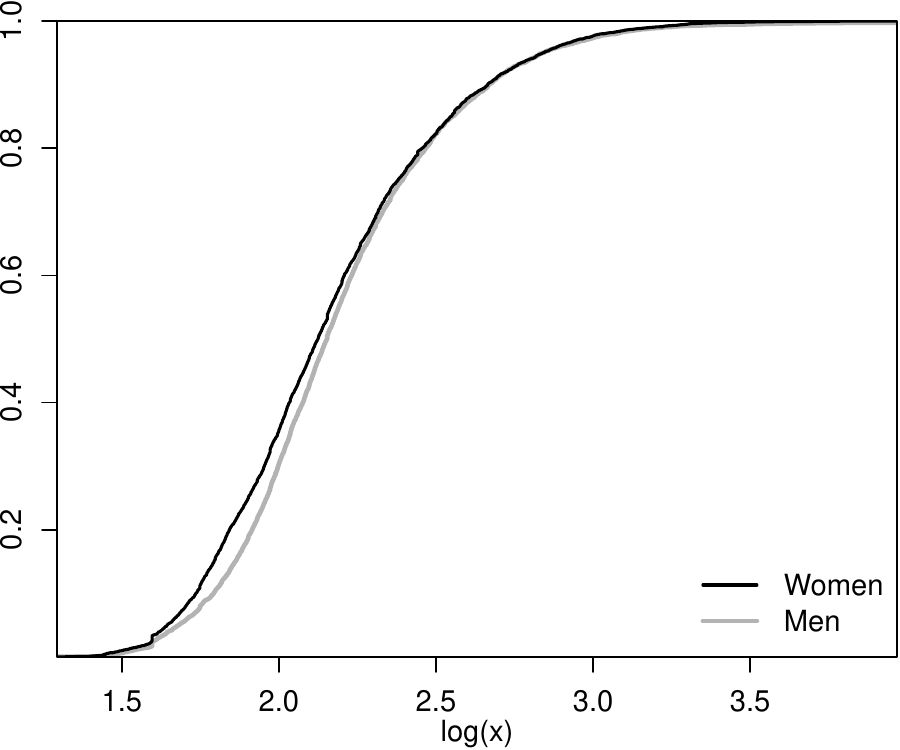} \\
    (a) & (b)
    \end{tabular}
    \caption{Ecdf's of the hourly salary \(X\) of a (a) Spanish worker and (b) full-time Spanish worker between 20 and 29 years old. Wages are in a natural logarithm scale.}
    \label{fig:ECDFsSALHOUR2}
\end{figure}

In this context the ASD test (\ref{Test.MVR}a) is very useful. The MVR \(\epsilon_0\) in \eqref{MVR} provides a quantitative measure of the ASD degree and, consequently, of the gender gap. A MVR close to \(0\) in the test of almost dominance of male wages over female ones (\(H_1\)) is an objective answer supporting the conclusion of gender inequality.
To illustrate the procedure we use microdata from the 2018 Wage Structure Survey in Spain, available at the web page of the Spanish INE (Instituto Nacional de Estad\'{\i}stica).
The variable under consideration is the salary per working hour; alternatively, we also examined the annual gross salary. There are many factors that can be taken into account: full- and part-time contracts, age cohorts,\dots{} For the majority of these groups, the empirical distribution function of men's salaries is clearly below or equal that of women, and generally there is a huge difference between both functions except in the lowest percentiles. Figure~\ref{fig:ECDFsSALHOUR2} is an example of this situation: we display the distribution functions of female (\(n_1=94168\)) and male (\(n_2=122558\)) hourly salaries. The men's cdf is less than the women's for all the sample values. The empirical distributions are ordered according to the first order and, thus, the associated empirical 2DSD index lies on the diagonal (see Suppl. Figure~3a): \( \hat{\mathcal{I}}_F (X_1,X_2) =(2.4218,2.4218)\) and the estimated MVR \(\hat\epsilon_0\) is 0.
%But, what implications does this have regarding the first-order dominance of male salaries over women's at the population level?
For a significance level $\alpha=0.05$, the bootstrap ASD test (Case 1), with \(B=2000\) bootstrap samples, yields \( \hat\epsilon_0^{*(0.95)}=4\cdot 10^{-6}\). Thus, \textit{there is a strong statistical evidence} supporting that men's wages almost dominate women's in the usual order and, given the extremely small value of \( \hat\epsilon_0^{*(0.95)}\), that the range of salaries where strict domination fails is negligible in comparison with the rest.
%Regarding second-order stochastic dominance, the 2DSD index in this case lies on the diagonal \(L_1\): \( \hat{\mathcal{I}}_2 (X_1,X_2) = (2217.596,2217.596) \) and the value of \(\epsilon^*\) obtained with 2000 bootstrap samples is less than \(10^{-9}\). Thus, males wages almost dominate female salaries also in the second order (as we could expect since first-order domination implies second-order one).

Considering only full-time workers between 20 and 29 years old (\(n_1=6100\) and \(n_2=9249\)), we obtain an interesting situation where \(\hat F_2\) is generally below \(\hat F_1\), even though they cross in two occasions, in the left tail and near the 90th percentile, and are fairly close; see Figure~\ref{fig:ECDFsSALHOUR2}. As expected, the empirical 2DSD index for the usual stochastic order is very close to the diagonal: \( \hat{\mathcal{I}}_F (X_1,X_2) =(0.9850,0.9864)\) . For a level \(\alpha=0.05\) and \(B=2000\) bootstrap samples (Suppl. Figure~3b), the test (\ref{Test.MVR}a) with the alternative hypothesis that men's wages ASD women's in the first-SD yields \(\hat\epsilon_0^{*(0.95)}=0.047\) (Case 1). Hence the area between the cdf's where the strict dominance fails is, with 95\% confidence, less than 5\% of the overall area between them, which clearly indicates a high degree of first-order ASD of men's earnings over women's.
%With respect to the second order the situation is more extreme: the 2DSD index \( \hat{\mathcal{I}}_2 (X_1,X_2) = (734.5240,734.5242) \) and \(\epsilon^*<10^{-6}\) (Figure~\ref{fig:RejectRegALDSALHOUR}(b)) indicate a marked second-order ASD of male salaries over female ones.

To illustrate the performance of the ASD test described in Case 2 of Section~\ref{Section.ASDtest} we selected the hourly salaries of individuals employed in the public sector, whose empirical distribution functions for the two sex apparently coincide on the interval of lowest wages (Supplementary Figure 4). There are \(n_1=19719\) women and \(n_2=15834\) men. For a significance level \( \alpha=0.05 \), \(B=2000\) bootstrap samples, \(a_n = c \, \log(n_1n_2/(n_1+n_2))/r_n\), with \(r_n=\sqrt{n_1n_2/(n_1+n_2)}\), we carried out the test \eqref{Test.MVR}(a) on a grid of \(\epsilon\)'s and for different choices of \(c\). For each \(c\), we determined the minimum \(\epsilon\) for which \(H_0\) is rejected (Supplementary Table 1). These values of \(\epsilon\) are all less than 0.0016, which again supports the conclusion that, even in the public sector, stochastic dominance of male wages fails to hold in a very small interval of salaries.
\color{black}

\subsection{Almost Lorenz dominance between income distributions} \label{Section.ALDincome}

When the variable \(X\) is the income of a household, the distribution
%of income in the surveyed population
is frequently described by the Lorenz curve.
%As mentioned in Section~\ref{Section.index},
The MVR \(\epsilon_0\) and the 2DSD index \(\mathcal{I}_\ell (X_1,X_2)\) associated to the Lorenz order measure the relative inequality between two income distributions, \(X_1\) and \(X_2\).
%The parameter \(\epsilon_0\) in \eqref{MVR.Characterization} is 0 if and only if \(\mathcal{I}_\ell (X_1,X_2) \) lies on the diagonal \(L_1\) (see Proposition~\ref{Proposition.I.properties.1}) and this is equivalent to the strict ordering \(X_1\le_{\text{L}}X_2\), that is, income distribution being strictly more unequal in the second population than in the first one.
We consider the ASD test (\ref{Test.MVR}a), with respect to the Lorenz order, to check whether income distribution in population 2 is almost more unequal than in population~1 (alternative hypothesis).
Using income microdata from Spain, available at the INE database, we determine the equivalised disposable income (the variable \(X\)).
%, used by Eurostat to compute %the Gini index and other inequality descriptors among the populations of the European Union.
%, that is, the total disposable household income divided by the equivalised household size.
%The total disposable household income is the total income of a household that is available for spending or saving, after incorporating the net interest and dividends received and the payment of taxes and social contributions.
%The equivalised household size is the number of household members converted into equivalised adults by the modified OECD (Organisation for Economic Co-operation and Development) equivalence scale.

We observe that the ASD test can be used to perform a cross-temporal comparison of a random variable holding fixed the rest of conditions, for instance, the location (Spain in this case).
Population 1 is then the variable of interest (Spanish income) in a reference year (here 2008).
Population 2 is the same variable in a posterior year.
The test (\ref{Test.MVR}a) checks the almost Lorenz dominance of the income \(X_2\) over the income \(X_1\) in 2008.
The year 2008 was chosen as reference because it marked the start of a deep financial and economic crisis in Spain and several other countries. Although the recession that took place in Spain ended in 2014, income distribution in later years is still more unequal than in 2008.
The estimated 2DSD index and the quantile \(\hat\epsilon_0^{*(0.95)}\) derived from the test reflect the changes in income inequality between the two years; see also \cite{Baillo-Carcamo-Mora-2022}.

\begin{figure}
\centering
\begin{tabular}{cc}
\includegraphics[width=0.51\textwidth]{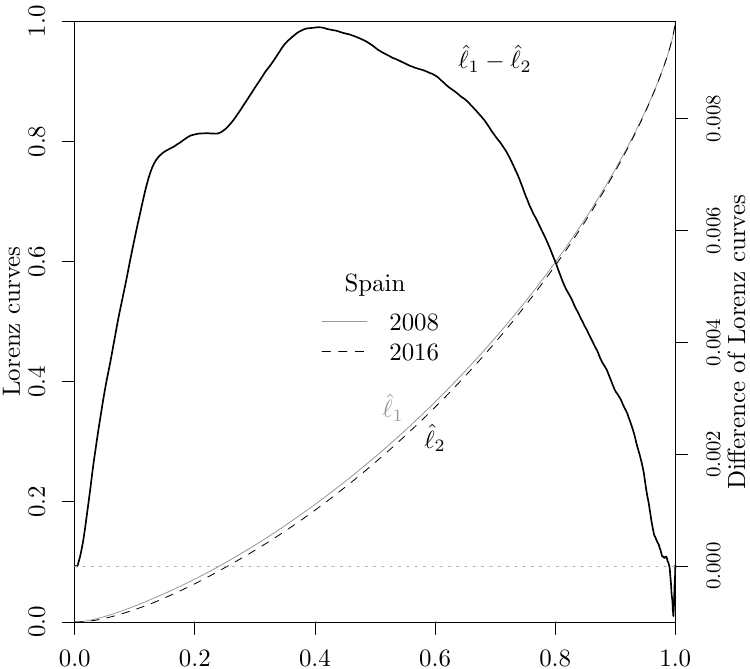}
&
\includegraphics[width=0.45\textwidth]{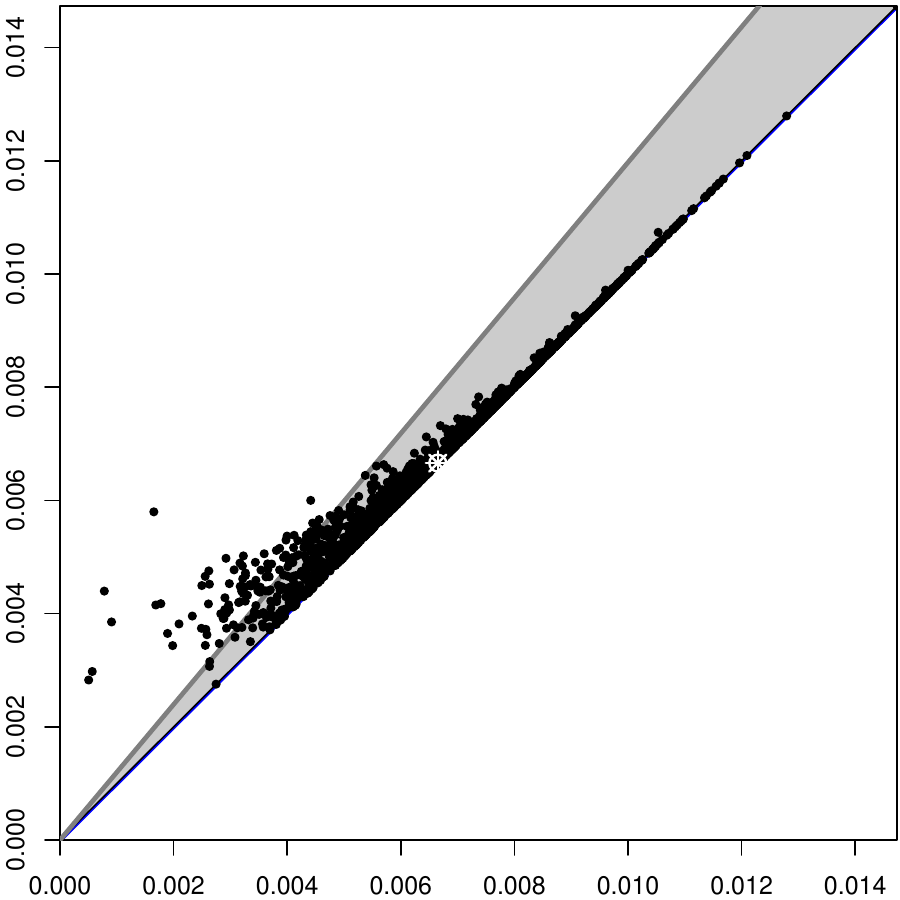} \\
(a) & (b)
\end{tabular}
\caption{Spanish income in 2008 and 2016: (a) Lorenz curves and their difference; (b) index \(\hat\I_\ell\) (white point), bootstrapped indices \(\hat{\mathcal I}^{*b}\) (black points) and \(R_{1,\hat\epsilon_0^{*(0.95)}}\) (grey sector).}
\label{Fig.Spain0816Lorenz}
\end{figure}

\begin{figure}
\begin{center}
\includegraphics[width=0.7\textwidth]{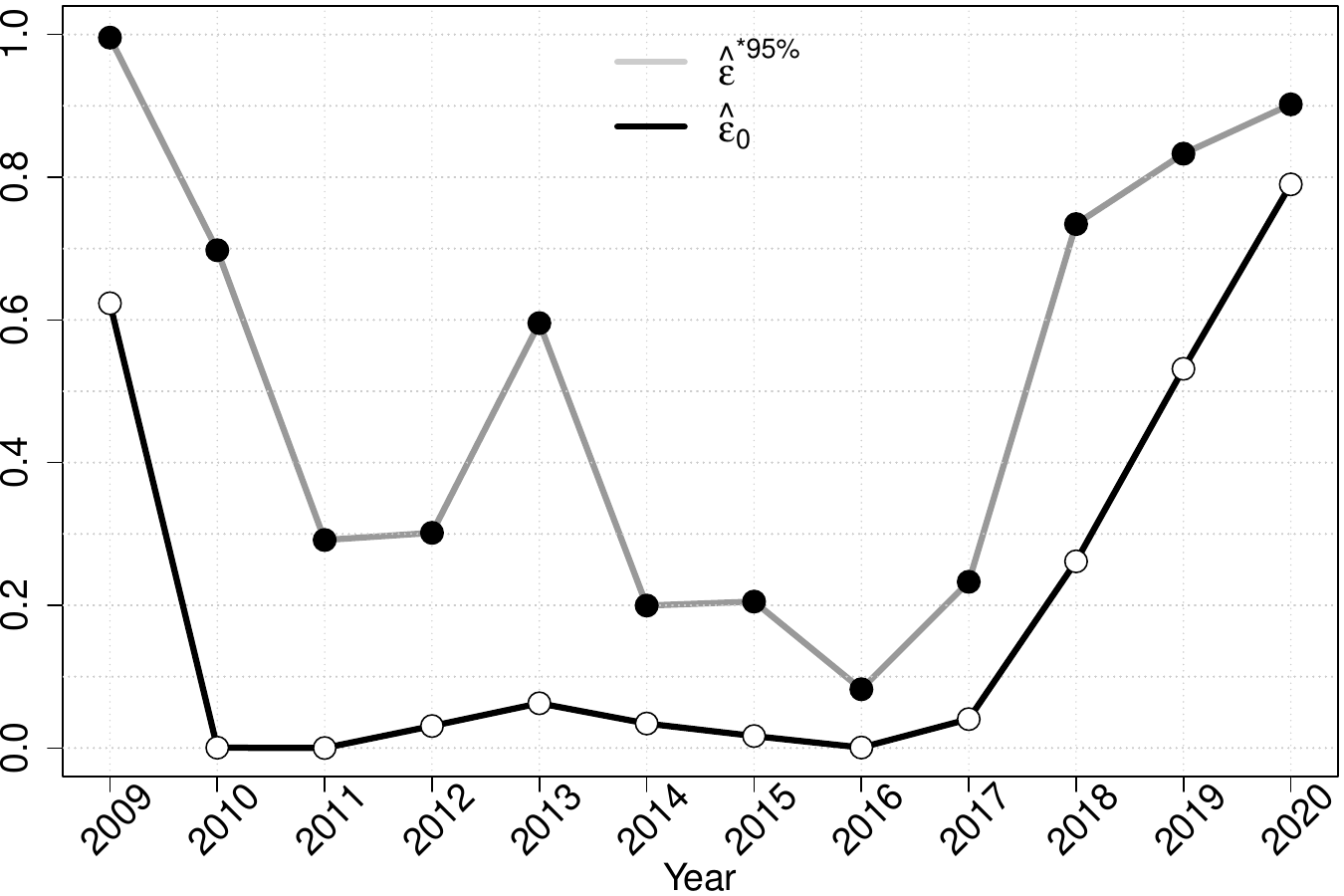}
\end{center}
\caption{Evolution of \(\hat\epsilon_0\) and \(\hat\epsilon_0^{*(0.95)}\) in Spain for 2008 and each year in the span 2009--2020.}
\label{Fig.Spainepsilon}
\end{figure}

Depending on the year posterior to 2008, the result of the ASD test is actually very different.
%We illustrate this with 2016 and 2018.
For instance, when \(X_2\) is the income in 2016 (see Figure~\ref{Fig.Spain0816Lorenz}(a)), the Lorenz curve of 2008 is above or equals that of 2016 for every percentile except an interval contained in $(0.99,1)$. The situation is a clear candidate for Lorenz ASD. Indeed, the estimated 2DSD index (Figure~\ref{Fig.Spain0816Lorenz}(b)) is \(\hat\I_\ell = (0.006773,0.006777)\) and the ASD test carried out with \(B=2000\) bootstrap samples gives \(\hat\epsilon_0^{*(0.95)}=0.082\), with $\alpha=0.05$. In other words, in 2016 the income distribution in Spain was almost more unequal than in 2008. However, in 2018 the economic and financial situation in Spain had substantially improved. The Lorenz curve of 2018 is above that of 2008 in the interval (0.68,1). The value of \(\hat\epsilon_0^{*(0.95)}=0.73\) (see Supplementary Figure 5) indicates there is no almost dominance in this case.
In Figure~\ref{Fig.Spainepsilon} we display the estimated MVR and the minimum \(\epsilon\) for which there is evidence that a year almost Lorenz dominates 2008. The (almost) increase in income inequality with respect to 2008 is evident for 2011, 2012 and from 2014 to 2017.
Note that, even if \(\hat\epsilon_0\) is very small in 2010, the Lorenz curves of 2008 and 2010 are so near each other that the estimated index \(\hat\I_\ell\) is very close to the origin and the bootstrapped indices are distributed all around the origin in the grey region of Figure~\ref{fi:propiedades} (see also Supplementary Figure 5). This explains why the \(\hat\epsilon_0^{*(0.95)}\) corresponding to 2010 is so large compared to the estimated MVR.

%%%%%%%%%%%%%%%%%%%%%%%%%%%%%%%%%%%%%%%%%%%%%%%%%%%%%%%%%%%%%%%%%%%%%%%%%%%%%%%%%%%
% SECTION: ACKNOWLEDGEMENTS
%%%%%%%%%%%%%%%%%%%%%%%%%%%%%%%%%%%%%%%%%%%%%%%%%%%%%%%%%%%%%%%%%%%%%%%%%%%%%%%%%%%

\section*{Acknowledgements}

We are very grateful to the Editor, Associate Editor and two reviewers for all the comments on the first version of the paper. We especially appreciate that we were encouraged to consider the case where the contact set has positive measure, which resulted in Theorem \ref{Theorem.Modified.bootstrap}.
%The Associate Editor also pointed us to the reference \cite{Linton-Song-Whang-2010}.

\section*{Funding}

Supported by the Spanish Agencia Estatal de Investigaci\'{o}n
through projects PID2019-109387GB-I00 (A.B. and J.C.), PID2021-124195NB-C32 (C.M-C.) and the Severo Ochoa Programme CEX2019-000904-S (C.M.-C.).
C.M-C.\ has also been supported by the ERC Advanced Grant 834728.
%The authors report there are no competing interests to declare.
%This paper is based on data from
%the Instituto Nacional de Estad\'{\i}stica (Spain).
%The responsibility for all conclusions drawn from the data lies entirely with the authors.

%%%%%%%%%%%%%%%%%%%%%%%%%%%%%%%%%%%%%%%%%%%%%%%%%%%%%%%%%%%%%%%%%%%%%%%%%%%%%%%%%%%
% BIBLIOGRAPHY
%%%%%%%%%%%%%%%%%%%%%%%%%%%%%%%%%%%%%%%%%%%%%%%%%%%%%%%%%%%%%%%%%%%%%%%%%%%%%%%%%%%
%\newpage

\end{document}